Edith Cowan University

School of Communications and Arts

Faculty of Education and Arts

# Reimagining the Living Room

Andrei Petrut

10353517

i24 Master of Professional Communications

Supervisor: Keith Smith

# Table of Contents:

**Abstract**

**Acknowledgement**

**Abstract**

The following study is an attempt to seek out and understand emerging trends and some of the challenges we face when dealing with an increasing amount and complexity of entertainment technologies in our homes, particularly in spaces that are both private and public. Accepting the premise that this influx of technologies will continue, I set out to identify some of the challenges that consumers, manufacturers, distributors and creators of content are facing today in incorporating these technologies into home life. Anticipating a transformation, I tried to imagine what a modern family would expect from it. To address some of the issues uncovered in this research, I propose a new approach to smart homes and an updated model for content distribution, one which, I hope, would benefit all parties.

**Acknowledgement**

I wish to thank my family, my wife Raluca and my daughters, Emma and Julia. It's only thanks to them that I managed to pursue and complete this dissertation.

I am infinitely grateful to my collaborators, architects Raluca Ionescu and Ioana Tanasie. They took my pencil sketches and turned them into an intelligent and cohesive spaces. I owe many thanks to Travis Laidlaw, one of the most talented motion designer and visual effects artist that I had the pleasure to work with, for bringing life to this space.

I am also extremely thankful and indebted to my supervisor, Mr. Keith Smith, co-ordinator with the Film and Video School, who managed to steer me in the right direction and keep me positive and on track during this past year.

And finally a big thanks to everyone else who knowingly or inadvertently contributed to my presence here, on the beautiful West Australian coast or to my enrolment in this Masters of Communications program at Edith Cowan University.

Andrei Petrut

## 1. Introduction

The scope of this project is to redefine the living room based on current social and technological trends. I will identify the various ways in which people use this space, for formal and informal settings, entertainment or other social activities and use these findings as a starting point to reimagine the living room into a flexible, people-centric environment capable to adapt and cater to current and future expectations.

This model will incorporate architectural, design and technological elements in a unified and seamless concept, inspired by the users' expectations and focused not only on the range but also on the quality of their experiences. The integration of various gaming services with content delivery platforms for movies, television and interactive content will enhance these experiences and provide a simpler and user-friendly way to access this type of content. This integration is a critical aspect in achieving a higher standard of home entertainment, due to the sheer volume of content available, the wide range of originating sources and perceived or implicit relevance to the consumer.

A second component of this thesis is a short video featuring some aspects of the proposed model The video can be found here: https://youtu.be/I4GamoOo6sA or as a hard copy (usb or DVD), attached to the final submission.

This model will largely rely on existing technologies but since, at the present time, there is no initiative to integrate them into a unified platform, I will argue the need for adopting universal and open standards. In turn, this will require integrating the service constellation approach into current business practices, which shifts the focus from protecting ideas and technologies to improving user experience and accelerates adoption of new technologies.

Is there a demand for such concept? I think there is. A study commissioned by Logitech,  a manufacturer of one touch remote controls among others, found that "virtually nine out of 10 […] people believe their home-entertainment experience would be more enjoyable if they could just push a single button to enjoy their favourite feature film" (Logitech, 2010). Over time, our living rooms have acquired a number of technologies which in turn have shaped the experiences that we associate with it. I would like to explore the idea that these technologies can be harmoniously integrated into this living space expanding the range and enhancing the quality of our entertainment and social experiences.

Three key questions have shaped this research and are treated as separate chapters which address them. They are:

Question #1 - How has the living room evolved as a social and entertainment space in contemporary Western culture - how can this evolution be mapped in terms of changes in social activities, physical layout (design) and the utilization of screen entertainments?

Question #2 - What predictions have other researchers/agencies made about the future living room and screen entertainments - how might the nature of it change as an entertainment space and for what reasons?

Question #3 - What alternative and/or augmented model for the future living room and screen entertainments could be proposed - what legal and economic barriers might resist this and how could they be overcome in an alternative model?

## 2. Looking back

Question #1 - How has the living room evolved as a social and entertainment space in contemporary Western culture - how can this evolution be mapped in terms of changes in social activities, physical layout (design) and the utilization of screen entertainments?

### 2.1 Why the living room?

I chose to focus my research around this space for a number of different reasons. First of all, it is a room present in most households, and it is the space that will most likely be associated with entertainment and social activities. Secondly, as a common space in the household, I feel that it represents all the members of the family, present and sometimes past.

The consensus among researchers is that, in most dwellings, there is a designated room or space for public activities. Rechavi's research details the people's uses and experiences of their living rooms. She mentions studies that span across the globe, from Native America and Peru to Alger, Egypt, Israel and Japan. These studies show not only that virtually all cultures developed the concept of hosting and socializing but they also imply a universal distinction between public and private spaces. The same concept is found in 17th century French Big House, whose occupants would sleep in the chamber, while a separate area was reserved for social activities and hosting. This distinction spread throughout Europe and in the late 19th century, parlour and sitting room were attributed to a "room used by families of different classes for their common social activities and entertainment of guests" (Cromley, 1990). The term living room, however, was used to describe a space that had a similar function but was also used by the household members in the absence of guests. Rechavi suggests that the "middle-class supposedly began to use all the rooms that were designated for entertaining for its own leisure" (Rechavi, 2004). It is quite likely this was a direct consequence of industrialization which changed the place where men traditionally worked, from home to factories and other workplaces outside of home. Thus, the social role of the home changed, because "home not only sheltered the family, it helped bring the family together, and living rooms, as rooms where all members of the household could come together, were where such mending took place" (Rechavi, 2004). At the turn of the century one could still find a parlour, a sitting room, a drawing room or any combination of the above, all lumped together as living rooms, but the growing middle-class was about to change all that.

## 2.2. Family and the living room

Rechavi argues that, as the 20th century unfolded, a number of factors, such as the increasing percentage of funds allocated to technology like the automobile, the decrease in servant workforce, a reduction in the average number of children and a rapid urban growth, led to smaller apartments, which needed a multifunctional room to accommodate private and public aspects of the inhabitants' lives (Rechavi, 2009). The mass production of furniture also facilitated the introduction of the sofa, or "the Davenport" as it was called at the time, and Rechavi notes that:

> the connection of homes to gas and electricity in effect facilitated a new interior arrangement of living rooms, whereby the center table with its lamp, no longer needed in the center, moved towards the couch. This change coupled with an increased interest in domesticity and the accessibility of plush and well-designed couches for the middle-class turned the couch into a visual focal point, as well as a center of activity in the living room. (Rechavi, 2008)

The importance of a seating area is noted by most of Rechavi's interviewees, sixteen middle-class residents of the Metropolitan New York area, between 29 and 57 years old, who stated that "the one object without the living room would cease to exist as such was the couch" (Rechavi, 2009).

There was a slight correction to this downsizing trend, when in the 1950s and 1960s the prosperity brought about by the end of the war can be clearly seen in the plans of the new homes. In addition to the larger size, we find a new room, the family room, which Rechavi thinks is a direct consequence of "a growing interest in family life". She quotes Halle (1993), conceding that "family rooms or "dens" were intended to counter forces that pulled the family apart". In reality, the family room provided families with a more private and casual space that duplicated the attributes and functions of the living room.

A possible limitation of Rechavi's own research is the relatively narrow scope, being concentrated in New York City and as a result, most of the participants' dwellings are apartments. I still consider this data to be relevant for two main reasons. First, there is a global trend for urbanization; according to the World Bank group, somewhere in 2007 the urban population has officially surpassed the rural population and it is estimated that by 2025, the number of people living in metropolitan areas will reach 70 percent of the world's population (World Bank Group, 2015). Secondly, New York is the most populous and influential city in the modern western civilization. New York is also part of the largest entertainment market in the world (by box office). Some research even describes New York as the "Media Capital of the World" with an estimated entertainment and media spending of 19.7 Billion dollars in 2014 (statista.com).

Rechavi's own literature review, suggested that the living room, as the most public private space, is used for display purposes. While this might imply that it might be a space void of any personal importance, Rechavi found that there is a very strong link between people and their living rooms. Most participants had objects on display that were important and held a great deal of significance for them. It simply becomes a matter of how much the host will reveal about a particular object and its meaning.

For the purpose of designing this space and defining some of its characteristics, I am going to look at Rechavi's findings in a rather simplistic manner. Some activities may be more important than others, through frequency, number of participants, etc. I will not prioritize and attempt to give equal consideration to all, because Rechavi's research suggests that participants experience and relate differently to a certain activity and its outcome.

In no particular order this space is used for: entertaining, watching TV, listening to music, talking on the phone, working, studying, reading, eating, meditation, relaxation, contemplation, conversation, exercising, having sex, napping and occasional sleepovers. One participant mentioned that his living room was used regularly as a bedroom. but he added that he intended to physically divide the space, suggesting that this arrangement was temporary and fortuitous. As a result, regular, overnight sleeping has not been included on my list. In addition to facilitating these activities, for singles, couples and groups, the living room has to accommodate furniture, photographs and other objects of interest, mementos from certain places or times that held certain significance for their owners (Rechavi, 2004).

With the exception of one adult child, there were no children present in the dwellings sampled. Finding pertinent, up-to-date research about children and teenagers in this space has been challenging. We know they are watching live TV and that they are playing games on computers and gaming consoles. To what extent these activities take place in the living room is unclear, because, according to Henry J. Kaiser Family Foundation, 71% of young people, 8-18 year old, had a TV set in their bedroom and 50% a video game console (2010). We also know that some of them use computers and other devices (mobile phones, tablets and iPods) for texting and social media activities: "7th-12th graders report spending an average of 1:35 a day sending or receiving texts" and 74% of them "say they have a profile on a social networking site" (Henry J. Kaiser Family Foundation, 2010).

A more recent study, commissioned by the Entertainment Software Association, claims that "51% of U.S. households own a dedicated gaming console and those that do, own an average of 2" (2014). Given that 18% of al gamers play with their parents and that 14% play with their spouse or significant other, I think it would be safe to assume that in some households the gaming console is located in a neutral area, such as the living room. The ESA report mentions that the number one reason for parents playing games with their children is that "it's fun for the entire family" (ESA, 2014). This would also be consistent with my personal experience where members of the Y generation, grew up with this technology in their bedrooms because the main TV set was reserved for live broadcasts and cable TV. Now, as home owners, they are perfectly happy to relocate the gaming console into their living rooms. For all these reasons, I will add gaming to the list of the activities taking place in our living room.

Rechavi splits these activities in three ways: "one type of activity is getting together with people who are not members of the household; the other is […] "being with oneself." The third is spending time with people who are part of the household" (2004). One must assume that in order for this to happen some compromises which affect the functionality of this space must be made. Also, certain activities might be dependant on the same technology, i.e. watching TV and playing games, so certain time compromises must be made as well. In some households, these compromises have been mitigated by adding another room altogether, the family room. In my personal experience, this room is similar in purpose to the living room and the family member's preference for one or the other is simply arbitrarily. It does seem to cater a bit better for the needs of a younger generation, possibly due to its less formal atmosphere. In a situation where multiple guests would be present, the older generations would prefer the living room while the younger, less formal generations would feel more comfortable in the family room. However, the homes with family rooms are certainly a minority and I think will continue to be, in light of the global trends for urbanization and population growth.

### 2.3. Entertainment and Communication technologies in the living room.

The early technologies that we could find in the living rooms, even before the turn of the century, would be phonographs or record players. In the homes of the more affluent these became a status symbol and provided entertainment to their guests. Despite shortcomings, such as low volume, poor sound quality and a limited selection of content, their convenience, being able to play music without any previous arrangements, was enough to guarantee success.

Hot on the heels of the telegraph, radio was the first true broadcast medium and almost immediately, "played a vital role in the formation and maintenance of national identity" (Giblett, 2008). Often combined with entertainment, music and soap operas, its crucial role in polarizing nations was felt all over the world. In the United States, it "brought the nation together as a single, vast audience; one that could be addressed intimately, yet simultaneously" (Hanson, 1978, as cited in Giblett, 2008). The uptake of this technology by the middle class in the 1920s, established Commercial Radio as one of the most successful enterprises of its time. Tucker (1978), claims that, in 1922 there were "an estimated million listeners and nearly 600 broadcasting stations in the United States" (Giblett, 2008). The impact though, was being felt everywhere and Rechavi (2008) notes:

> The decrease in size, the improved design and the collaboration with the furniture industry was intended to bring the radio into the living room and give it at least the same place of honor as the phonograph had. Once the radio did become more agreeable in appearance and even attractive on its own, it not only entered the living room, it became a visual focus and an activity center for the household. The era of the "entertainment center" had arrived.

Radio came into our lives with minimal disruption to the activities taking place in a home and in many ways, complimented and enriched our lives. However, it wasn't long until radio became a household necessity. Hendy explains: "In America, […] coverage of the deepening crisis in Europe in 1938 saw over two-thirds of those Americans polled claim radio as their preferred news source (Scannell, 1990 as cited in Hendy, 2003). At the same time the movie theatres were running newsreels before and in between main features and some even featured newsreels exclusively. Although television was successfully demonstrated in 1939 at the World's Fair in New York, a lack of agreed standards and then America's entering the war delayed its widespread adoption. During the war, only Nazis continue to broadcast television, a testament to the power and technological advancement of Germany.

After the war there was a short period when television seemed to take off in the US, and according to Winston, "there were four networks, fifty-two stations and nearly a million sets in twenty-nine cities" (Winston, 2003). But a four year freeze by FCC (Federal Communications Commission) was going to be the last obstacle for television. The main issues were: a standard for colour television, the reservation of channels for noncommercial and educational television, reduction of channel interference, the lack of a national channel allocation map and finally, the opening up of new spectrum space. After the freeze, which ended in 1952, television skyrocketed: "the number of TV stations jumped to 573, broadcasting to nearly 33 million receivers" (Winston, 2003). And this was a worldwide phenomenon: "By 1970 there were 231 million sets in the world [...] and the system laid down a decade earlier was stable" (Winston, 2003).

All of this took place despite early concerns that television would disrupt the American home and family. In 1948, a New York Times TV critic observes that "the wife scarcely knows where the kitchen is, let alone her place in it. Junior scorns the late-afternoon sunlight for the glimmer of the dark living room. Father's briefcase lies unopened in the foyer. The reason is television" (as quoted by William Boddy, 1998). Furthermore, William quotes a Parent's Magazine reader, who "described her family's successful adjustment of daily routines to accommodate television" and "complained of adult neighbors 'who insisted on conversing' during the evening's television entertainment" (William, 1998). The programming disrupting these American families consisted mostly of sports news, because half of the audience was watching television on one of the 3000 sets operating in bars (William, 1998). During the 1950s however, as middle income families acquired TV sets, new kinds of programming appeared, most notably situational comedies such as "I Love Lucy" which "was the biggest hit on the little screen" (Winston, 2003). This type of entertainment programming continued to fuel the appetite for television, already in full swing largely due to suburbanization and the baby boom (William, 1998).

Indeed, America was in the middle of a social revolution: "mass exodus to the suburbs, new realms of leisure, rising incomes and a tremendous demand, both for things and for entertainment", according to a 1956 Business Week article (William, 1998). And, in typical American fashion, television was going to be shaped by the moral and social values of this middle class. Pressured by Catholic groups, FCC commissioners and Congressional investigators, the industry established the Television code in 1951 (William, 1998). As one NBC executive suggested: "vulgarity, profanity, the sacrilegious in every form and immorality of every kind will have no place in television. All programs must be in good taste, unprejudiced and impartial" (as quoted in William, 1998). The other, less obvious censorship of this medium was hidden in its aesthetic mission: "The style of acting in television is determined by the conditions of reception; there is simply no place for the florid gesture, the overprojection of emotion, the exaggeration of voice or grimace or movement" claimed Gilbert Seldes (quoted in William, 1998). The true power of television, its "very lifeblood and magic" as CBS president Frank Stanton calls it, was recognized in its capacity for "live transmission of dramatic or unstaged events" coupled with the national networks' ability to distribute this service nationwide (William, 1998). At a time when the intercontinental nuclear missile crisis was shaping policies, a service capable of reaching nearly every home in America, at a moments notice, was allowed to be used for political purposes under the guise of national unity (William, 1998).

The Ford foundation was first to commission a content analysis of TV programming, which concluded that between 1949 and 1951 TV educational programs amounted to less than 1% of air time and advertising to roughly 20% (William, 1998). However, neither the first wave of educational television in the mid 1950s, the second one in the early 1960s or the Hollywood's

Pay TV, managed to disturb the growth of commercial television in the US and subsequently the rest of the world. The three major networks, CBS, NBC and ABC controlled the access to primetime viewers and leveraged that advantage in acquiring and syndicating independent productions to audiences all over the world. This had a profound influence on the quality of the programs because, as TV critic Robert Lewis Shayon remarked: "Such programs must not only be aimed at the lowest common denominator in this country; they must also be geared to the potential audiences of nations whose emergent cultures are largely at a primitive level" (William, 1998). More importantly, I think that somewhere in the late 50s and early 60s, commercial television established a tacit agreement with its audiences: we would watch, and subsequently fund the commercial stations through advertising and in turn, we would have access to free, quality programming.

And this is what television was shaping up to be, an always changing, rapidly evolving cutting edge technology which allowed the simultaneous distribution of audio and visual information, from struggling educational programming to profitable, mass-appealing entertainment and everything in between. It's important to make this distinction and acknowledge this duality early on, because these different aspects will influence and depend on each other throughout the evolution of television. As O'Sullivan remarks, "the television set was an 'invisible apparatus'. Memories initially focus much more on programs and the novel shared experiences of the early TV viewing" (O'Sullivan, 2007).

The technological evolution of the television set consisted in gradual increases in picture size and resolution, improved sound reproduction, remote control and various interfacing capabilities. What started as a 9 inch black and white image, with a few hundred lines of resolution and a narrow range monoaural sound, has come very close, in recent years, to matching our biological capabilities to perceive image and sound. It even challenged these boundaries by forcing 3D perspectives in spaces which are not three dimensional in nature. In addition, many television sets sold today, so called Smart TVs, can run various applications, such as streaming services, videoconferencing, even simple games. However, I think that this is a transitional period when this technology is redefining its boundaries and these overlaps with other technologies will disappear when the audio-visual capabilities will be "maxed out". This will also allow for other capabilities to be developed independent of the already matured characteristics.

Another technological aspect of the television has been its content delivery mechanisms. As a medium which evolved from radio, it started with a similar broadcasting system, but due to the use of a higher carrier frequency, the broadcast was limited by distance and geographical obstacles. These limitations were overcome by land based cable networks or geostationary satellite transmissions. In turn, this allowed for a much wider distribution of content and gave rise to an increasing number of specialized channels which could not survive in the traditional broadcast system.

During the 1970s the cable distributors began to separate themselves from broadcasters and argued that the 1934 Communications Act did not apply to them. According to Stadel "what was at stake in the debate was not merely the right of cable systems to distribute pornographic content, or the right of cable consumers to view pornographic films in the privacy of their homes, but a definition of cable as a new medium" (Stadel, 2014). Pornography was indeed the spark behind this issue because the highly regulated traditional broadcasters were unable to compete in this type of programming. This competitive advantage was also used to "justify the largely untested concept of getting American households to pay for television" (Stadel, 2014). Coopersmith, as cited in Stadel, argues that "sexually explicit material commands a higher price than other kinds of media because of both high consumer demand and the socially discouraged status of its consumption" (Stadel, 2014). The cable system also introduced a tiered model, leading to a fragmentation of the audiences, something that Stadel considers to be "a fundamental change in the medium" (Stadel, 2014). The "basic" tier was comprised of traditional free to air broadcasts, and it was complemented by the "premium" tier, which could be purchased on a "per-channel basis or in a bundle of such channels" (Stadel, 2014).

In allowing viewers to purchase a channel dedicated to sexual programming as a separate addition to their basic cable service, pornography helped form the basis of what would become the major distinguishing feature of cable, the selective and private nature of its audience" (Stadel, 2014)

I think this adoption of cable television is very important because it brought this content to the consumer, in exchange for a service fee. Also, it added a new dimension to our television sets transforming them into "a sexual technology that could be used as an adjunct to sexual practice, a kind of virtual sex toy" (Stadel, 2014).

Another technology that transformed our living rooms was the Video Home System, or VHS. Introduced in 1977, it changed the nature of television from "appointment television" where viewers scheduled their time according to their favourite shows, to a less rigid format, where viewers can record the content and see it later, at a more convenient time. I believe the seeds for the wide penetration of VHS were sown a lot earlier, in 1963, when Philips, along with the compact cassette, started selling the EL 3300, the worlds first cassette player and, more importantly, recorder. I think this recorder, along with the proliferation of 8mm amateur film cameras, radically changed how we perceived ourselves. All of a sudden, we were not only spectators, viewers and listeners, but we could also create and exercise control over content. We could star in our own movies, we could make our own music and we took over the role of the broadcasters and film studios, on a global scale.

When VHS arrived, a cheaper and easier to use alternative, we were ready and more than willing to pay for it. Even with the confusion associated with the Betamax/VHS format war, by 1988, "half of American homes owned VCRs" (Davis, 2004). An interesting feature of some VHS machines was the capability of "commercial skipping". The manufacturers introduced this feature in response to consumers fast forwarding through commercials. Even though broadcasters sued these manufacturers (Stoltz, 2012), this feature endured and it's still present on many DVRs, today's equivalent of a VHS machine.

I think there are two major drivers which governed the adoption of technologies in our living rooms over the years: control over content and ease-of-use/convenience. Every new technology that entered our living rooms was initially fuelled by our desire to access and control the content, a defining characteristic of the early adopters. In search for a better, more customized experience, they are willing to pay a premium in order to bring new technologies to market. They are also willing to experiment (early adopters of cable TV / pornography) and take certain risks, such as copyright infringement (commercial skipping, duplication). And then comes the bulk of consumers, looking for a one-size-fits-all, ready-made, proven solution, which makes their experiences a bit easier and/or better. Even though they are paying a lower price for these technologies, through sheer volume, they brings profits to all companies involved and thus, they are in fact financing the next wave of technological advancements. In stark contrast with the first group who are in fact living in future, always unhappy and looking for the next best thing, these late adopters are almost always using mainstream technologies on their way to becoming obsolete. These discrepancies, this desynchronization has fuelled and continues to drive technological advancement and the development of new technologies.

### 2.4. War in our living rooms.

For the past 70 years or so, the technological evolution born in the midst of the second world war had somewhat slow, predictable cycles. From black and white to colour television in 1965, approximately 20 years. From compact cassette to compact disc (CD) which arrived in 1981 we had a cycle of almost 20 years. From VHS to DVD (1995), again, almost 20 years. However, the CD and the DVD marked the passage from an analog format to its digital counterpart. For manufacturers, the adoption of these digital formats represented a cheaper and faster way to bring content to their consumers. In the analog system, duplication was a very time consuming operation, often performed in real-time, on expensive, industrial equipment. The digital technologies have all but eliminated the time component in this process. The time reduction and the cost savings go beyond duplication of content to manufacturing devices and equipment. The technological cycles have been reduced to a point where the hardware itself has become ubiquitous and irrelevant. This gave rise to a new type of consumer, one who simply does not see a device but an experience, one who lives not in the past or in the future. Mr. Right now.

In 1976, Fairchild Semiconductor introduced the Fairchild Channel F Video Entertaining System, the first gaming console based on a microprocessor architecture which used programable ROM cartridges to run video games. It turned the television from a passive medium into an active one. Using the screen as a medium, one could virtually change the game, i.e. the content, i.e. the experience, in 2 seconds flat. This was unprecedented, because now, one could decide the nature of the content, much like switching to a different TV channel, and also have control over the time, over characters and plot via a digital sublimation, an avatar. Perron and Wolf summed it up like this:

> The video game is unlike any media to come before it, the first to combine real-time game-play with a navigable, onscreen diegetic space; the first to feature avatars, and player-controlled surrogates that could influence onscreen events; and the first to require hand eye coordination skills. (Perron et al., 2003)

It started with very crude animations, depicting very simple tasks, like playing table tennis or shooting ducks and less than 30 years later we are creating entire online communities, virtual realities that can host and nurture the most intimate and complex human experiences. According to Perron et al., these massively multiplayer online role-playing games are "the first persistent (twenty-four hours a day, seven days a week) worlds, and the first instance of individualized mediated experiences within a mass audience (Perron et al., 2003).

When I think about this new kind of consumer, the one that lives right now, the one that creates his or her own reality, free of any time constraints, I see, for the most part, members of the Y generation, born between 1981 and 2000. One of their unique characteristics is that they were born into a world where computers have permeated all aspects of their society: they do not send letters, they send emails and text messages, they do not read maps but follow GPS instructions and they do not play board games but rather digital versions of them. They even fight wars in front of the computer screens.

The PBS documentary "Rise of the Drones" claims that "the Pentagon relies on a family of over 10,000 drones" and that "the Airforce predicts that nearly a third of its attack and fighter planes will be drones within a decade" (Yost, 2014). If we accept that most technologies in use today, have been developed or perfected in the war machine we should be expecting this kind of automation and virtual reality to find civil applications in the near future. We do see some signs of that; driverless cars seem to be on the books of most of the automobile manufacturers plus a number of other tech companies such as Google and Apple. Still, an unfulfilled reality today, but there is a legal framework in place to allow their testing on public roads (NHTSA, 2013).

If Airforce is controlling their unmanned airplanes from specialized facilities, because of obvious security concerns, when such technology finds its way into commercial applications, where will the control centre be placed? In some cases, services like a bus driver or an airplane pilot would require appropriate security measures, but not for a crane operator or a university lecturer, and such activities could be safely and efficiently performed remotely from home. The living room, with all the technology already present in there would be one of, if not the preferred choice for a controlling interface.

> … most researchers would agree that work and work processes are fundamentally transformed with the rise of mobile communication, and one of most notable change is the blurring of the boundary between work and the private sphere. While permanent connectivity allows work to spill over into homes and friendship networks, it is also likely that personal communication will penetrate the formal boundaries of work. (Castells et al., 2007)

However, in the last few years there have been a number of reports acknowledging security and privacy concerns associated with this connectivity, for example Smart TVs equipped with cameras and microphones (Ngak, 2013). Our mobile devices, smartphones and tablets have raised similar concerns (Clover, 2015 and Hern, 2015). Our cars can be hacked and controlled remotely (Yoshida, 2015) and our public transportation systems could be compromised leading to life-threatening circumstances (Westcott, 2015). Ironically, in one instance, a hacker claims to have controlled commercial airliners by hacking into their onboard entertainment systems (Perez, 2015).

I can only hope that we are becoming increasingly aware that online security and privacy are unrealistic expectations, even for federal government agencies (Liptak et al., 2015). The imminent arrival of the "Internet of Things", which promises remote control over thermostats, security systems, AC plugs, sprinklers, automatic pet feeders and smart scales (Merriman, 2015), will propagate this problem further into our lives, either forcing us to accept complete transparency or fuel a massive quest for real solutions to this issue. I can only hope that we still have a choice in this matter.

### 3. Right here, right now

Question #2 - What predictions have other researchers/agencies made about the future living room and screen entertainments - how might the nature of it change as an entertainment space and for what reasons?

### 3.1. "If it Ain't Broke, Fix It"

As James Wentling notes, "one is more likely to find the TV/entertainment centre next to the fireplace; often the fireplace is subordinated to the TV or media centre location" (Wentling, 1995). The statistics for Television penetration support this argument, and "ninety-eight percent of American households own at least one television set - likely the highest penetration rate for any single product or good in the world" (DuBravac, 2007). Wentling argues that this technology has significantly influenced the design of a dwelling, by "accommodating television, VCRs and sound systems into media rooms in larger homes, and in average homes - media walls are found in dens, living and family rooms" and he concludes that "the fireplace role as the "hearth" for gatherings has been taken over by electronic devices" (Wentling, 1995). His argument supports Rechavi's findings by describing the ideal design for a media wall, "made of cabinetry that can close and hide the television equipment when not in use, so that the room can be used for other purposes such as entertaining" (Wentling, 1995). The television set also facilitated newer forms of entertainment in our living rooms.

It's important to note that each new wave of technologies duplicated, at least some of the previous generation's functions. The radio would deliver news but also play music just like the phonograph. The television set itself is a radio with an added visual component. Initially, the gaming console served its purpose in a very purist way, yet with the current generation, always on, always connected 50% of gamers are also watching movies using their gaming console (ESA, 2014).

The newer generation of TVs can do all that, including gaming, albeit with certain limitations. Almost anywhere you look, there is an overlap of capabilities that has been inherited from generation to generation. Sometimes it's perceived positively, as a matter of choice and other times in a negative manner as a source of confusion, frustration or redundancy. A study commissioned by Logitech, a manufacturer of one touch remote controls among others, found that "virtually nine out of 10 (93 percent in America, 89 percent and higher globally) people believe their home-entertainment experience would be more enjoyable if they could just push a single button to enjoy their favourite feature film" (Logitech, 2010). I think this desire to simplify the way we interact with technology is the most important reason behind the widespread adoption of the smartphone.

It's the ultimate remote, the most versatile 5 inch surface that ever existed, the most accessible and rewarding piece of technology in our society. It can replace most if not all other devices we use to communicate, increase our productivity or to entertain ourselves. But most of all, for the younger generation, it's a social instrument, used to communicate faster and more efficiently than anything else. Quoting Oksman and Rautiainen (2002), Castells agrees that "the most important thing in mobile communication remains building up and maintaining their social networks" (Castells et al, 2007). Smartphones are currently, behind the internet, the second thing that Americans would not give up, and somewhere in the year 2015 the number of US households without a landline will be surpassed by the number of households which rely entirely on cellphones (Statista, 2015).

Whether to remain in contact with each other or to have access to online games and services, young people adopted mobile technology wholeheartedly. Ironically, this adoption was also driven by their age, which would prevent them from owning a landline. Also, the need for constant supervision on the guardians' side and the young people's desire for independence can be successfully mediated by this technology (Castells et al., 2007). Castells concludes that "wireless communication technology modifies but does not eliminate the power relations between parents and children" (Castells et al., 2007). "In some situations, texting is better than calling, not only because of the connotations of the communication channel, but because more time is expended in the activity itself" (Castells et al., 2007). Text messages allow the conversation to evolve over a longer period of time, without disturbing or interrupting other activities and in the same time provide a certain entertaining factor (Castells et al., 2007).

However, some compromises had to be made in order to make these devices portable. The relatively small screen, combined with our visual nature is the main reason why we still prefer larger displays. Neurologists claim that "upwards of 50% of the neural tissue is devoted to vision directly or indirectly" and that "two-thirds of the electrical activity of the brain is devoted to vision when the eyes are open"(Bowan, 2008). Also, the simple physics of the human eye will result in more strain when we focus on closer objects for long periods of time. The human vision is characterized by a central horizontal field of view of 50-60 degrees (information collected with both eyes) and approximately 55 degrees vertically. Our peripheral vision goes way beyond that, with more than 180 degrees horizontally and 120 degrees vertically (Bowen, 2008). Following the audio/visual reproduction standards set out by THX Ltd., which specify filling up approximately 36 degrees of vision (THX, 2015) when using a device with a 5 inch diagonal screen size, it should be placed 18.2 centimetres in front of your eyes, a rather uncomfortable distance for most of us. This makes the smartphone adequate for short, intensive, information loaded applications but prevents us from having a comfortable experience over prolonged applications. Thus, larger displays, like the TV screen still play a significant role in our lives. This is supported by the Office of Communications: "91% of UK adults view TV on the main set each week, up from 88% in 2002". This happens concurrently with the adoption of smartphones, "with over half of adults (51%) now owning these devices, almost double the proportion two years ago (27%)" and tablets: "ownership has more than doubled in the past year, rising from 11% of homes to 24%" (Office of Communications, 2013). It looks like we are not ready yet to throw away the TV set, but we might not need an antenna soon, because in its current form, television is about to be replaced.

### 3.2. What are we watching and how do we pay for it?

Traditional TV programming and its distribution channels are struggling to remain relevant. Over the top content providers, like HBO Now, at $15/month, are increasingly taking the place of traditional television providers. They commission or acquire exclusive content and this may seem to be a viable system for both producers and consumers of content, but in reality, once a show has aired or streamed, this competitive advantage disappears, and within hours, the content can be downloaded from various websites at no cost. I am not going to comment on any moral or legal issues that are associated with this practice, instead I will accept that it is a widespread phenomenon in our society: "more than $800 billion worth of content changed hands via illegal distribution networks in 2014", of which $104 billion is attributed to the movie, TV and music industry (Tru Optik Data Group, 2015).

Copyright protection has been a largely unsuccessful practice and the dissipation of digital technologies has made its mechanisms inefficient and obsolete. For example, on May 12, 2015, NineMSN was reporting that within 12 hours of streaming, the latest episode of "Game of Thrones", a popular HBO series, had been dowloaded illegally by more than 2.2 million I.P. addresses. It would be highly impractical for HBO to prosecute all these people and that process does not guarantee a positive outcome for them (NineMSN, 2015).

An interesting case study on the value of content could be done around the recent Mayweather, Jr. vs. Pacquiao boxing game. As New York Time highlights, the importance of this event can be immediately felt in the collaboration between HBO and Showtime, direct competitors in high profile sporting events. According to NYT, the suggested pay-per-view price was $89.95 to $99.95, "a record for boxing, a reminder of how the sport has marginalized itself by putting its best fights on pay-per-view" (New York Times, 2015). But that was the suggested price for the US market, with other networks broadcasting the event for free, in Columbia, China, Dubai, France and Mexico (Campbell, 2015). HBO and Showtime went to extreme lengths to protect this event from illegal downloads and obtained a court order against a few websites that were promoting the illegal streaming of this match. It seems the only way to achieve this was to ban online streaming altogether, with the court specifying that "there are no authorized online streams of the Coverage for delivery to United States audiences"
(United States District Court for the Central District of California, 2015). This resulted in "4.4 million PPV buys and over $410 million in domestic PPV revenue" (Lincoln, 2015). What the court order could not prevent was the use of streaming apps like Periscope and Meerkat. I could not find any numbers for Meerkat, but Periscope is owned and backed by Twitter, which reported a staggering 2.7 billion views on fight related tweets. And that certainly did not go unnoticed:

> If Twitter CEO Dick Costolo understood the implications of this activity, he sure didn't show it in a tweet that declared Periscope the "winner" of the night. There's no question the app got tremendous exposure that will build nicely off the 1 million downloads impressively achieved in just its first 10 days (Wallenstein, 2015)

In an ironic turn of events, that very morning, HBO was using Periscope, streaming live images from Manny Pacquiao's dressing room and using their Twitter account, to invite followers to watch this stream. This pretty much sums up the state of confusion in which this industry finds itself today: on one hand HBO recognizes the power and reach this new medium has, but because they have not yet figured out a way to secure earnings from it, they block its use in the actual event.

Around the same time, Warner Music Group was posting its earnings for the second quarter of 2015 and Stephen Cooper told investors that "for the first time, the company earned more revenue from streaming music services than from digital downloads." (Clover, 2015). According to the earnings call, Warner Music Group also "expects streaming growth will continue, and it believes that declines in download revenue will be a continuing trend" (Clover, 2015). In a digital environment, and more specifically when we think of streaming, the music industry has one fundamental advantage over film or television, because of the smaller data rate needed to deliver its content. For video streaming with its Flash Media Server, Adobe recommends a 1280x720 pixels, 2400 Kbps audio and video stream, where the video portion of the stream requires between 17 times and 36 times more bandwidth that the audio portion (Adobe, 2015). This ratio is maintained across other streaming platforms. When you take into consideration that downstream traffic from Netflix alone, in peak periods reaches 35% of all internet traffic in North America (Spangler, 2014), available bandwidth becomes a very real and significant bottleneck for streaming video. However, this barrier is quickly disappearing in many geographical locations, catching traditional content providers and distributors unprepared and hesitant.

In Convergence Culture; Where Old and New Media Collide, Jenkings describes this change:

> If old consumers were assumed to be passive, the new consumers are active. If old consumers were predictable and stayed where you told them to stay, then new consumers are migratory, showing a declining loyalty to networks or media. If old consumers were isolated individuals, the new consumers are more socially connected. If the work of media consumers was once silent and invisible, the new consumers are now noisy and public. Media producers are responding to these newly empowered consumers in contradictory ways, sometimes encouraging change, sometimes resisting what they see as renegade behavior. And consumers, in turn, are perplexed by what they see as mixed signals about how much and what kinds of participation they can enjoy.
>
> As they undergo this transition, the media companies are not behaving in a monolithic fashion; often, different divisions of the same company are pursuing radically different strategies, reflecting their uncertainty about how to proceed. (Jenkins, 2008)

I think this uncertainty invites the consumer to take charge. Today, we have more choices than ever and we have fully embraced the concept of "instant gratification". Furthermore, we have developed a fairly accurate sense of what we, as consumers are worth to the other party. One could argue that we were taught: from the excessive amounts that companies are willing to spend on broadcasting a 30 second ad during SuperBowl, US $4.2 million in 2014 (Kline, 2015), to the millions of apps available for smartphones. It is a common practice in every App store to have 2 versions of the same app side by side. One would be free and support its developer through

advertising and the other, "the PRO version", would be ad free but cost money. Some of the more expensive (over 99c) "PRO versions" may have additional features built in, but most apps are identical in functionality. I think the system works very well in this context and allows consumers to spend their money in a fair way (i.e. download the free, advert loaded app and try it out before purchasing the "PRO version"). Also, "consumers are willing to see in-app advertising in exchange for free content" (Gordon quoted in Harper, 2013). For many years, television employed a similar concept: free-to-air local programming alongside cable or satellite TV, which offered additional choices and specialty channels for a fee. In the absence of choice or its illusion, the consumer feels coerced, cheated and tries to gain control by circumventing the legal framework altogether.

The cost of going to a movie theatre might also help consumers justify this behaviour. According to Reader's Digest's website, we pay $2 on a 5c bottle of water and upwards of $5 for a bag of popcorn which costs 37 cents (Reader's Digest, 2015). Add to that the glamour and excesses associated with Hollywood, and piracy, as it applies to screen content, seems to be justified, at least to some extent. Because if you do download an illegal movie, you are only partly to blame for the loss of income suffered by its stockholders. Someone else has made the first copy and some else made it available online. This is how the torrent phenomenon took off, one of the more popular peer-to-peer file sharing formats, born in the aftermath of Napster's demise. Typically, peer-to-peer file sharing refers to exchanges of data directly between users, without the need for a central server. It's a very efficient and cost effective service, with users providing the hardware, the bandwidth and some of the administrative tasks such as troubleshooting, upgrades, etc. Torrent users download an index file which contains metadata, information required to find a specific set of files. Because of the relatively small size of a torrent file, the servers used to host these files can be easily duplicated and brought online as needed. This implies a certain level of coordination among the parties involved and access to resources needed to keep these servers online.

According to Cuevas et al., the majority of the traffic associated with torrent transfers, approximately 67%, was attributed to a very small number of publishers, around 100, which can be split into 3 three different categories: profit-driven publishers, fake publishers and altruistic top publishers (Cuevas et al., 2010). The group most invested in keeping the torrent repositories online would be the profit-driven publishers which seem to be responsible for "roughly 30% of the content and 40% of the downloads" (Cuevas et al., 2010). These individuals and/or companies use the torrent portals and files to advertise their own websites which are selling various services, products or feature extensive advertising. The second group, the fake publishers, are responsible for 30% of the content and a remarkable 25% of the downloads, when you consider that "portals actively monitor the torrents and immediately remove the content identified as fake" (Cuevas et al., 2010). The authors make a very interesting observation about this group:

> Fake publishers primarily focus on Videos (recent movies and TV shows) and Software content. This supports our earlier observation that these publishers consist of antipiracy agencies and malicious users because the former group publishes a fake version of recent movies while the latter provides software that contains malware. (Cuevas et al., 2010)

The third group, altruistic publishers, are responsible for approximately 11.5% of content and 11.5% of downloads. According to Cuevas, "many of these users publish small music and e-book files that require smaller amount of seeding resources" (2010). Their content seems to be better categorized and lacks any evidence of monetization. I think this group could be regarded as the modern day substitute for borrowing an LP, a VHS or audio cassette. The digital era and increasing computer literacy has made piracy a very convenient way to access content. More importantly, unlike its legal alternatives, piracy has no regional restrictions, where content may not be available to all audiences. The only prerequisite is a computer and access to internet.

An even more user friendly way to access copyrighted material is the "the world's most popular online video community" (Scoop Media, 2009). A quick search on Youtube yields many results for full length HD quality movies like "Mr. Bean's Vacation" (https://www.youtube.com/watch?

v=vFJ0WtiE3Yc) which in less than a month, (uploaded on April 25th, 2015 and assessed on May 22nd) was viewed 685,000 times. Another example would be "The Waterboy" (https://www.youtube.com/watch?v=Hur-HMbuTB4) uploaded on March 13 with more than 592,000 views by June 15th. And this happens quite regularly, despite Youtube's Content ID system.

Content ID is a system that uses preexisting profiling for content and looks for a match within the user uploaded content. For this system to work as intended copyright holders need to provide Youtube a copy of their work and Youtube needs to analyze and profile that content. If a match is found, Content ID gives the copyright holders the obvious choices, such as blocking, muting (in case of an audio only infringement), but also offers the option to "monetize". It basically allows the use of copyrighted material but injects advertising in the video or presents ads alongside. The proceeds resulting from these ads are redirected towards the copyright owners. This is a new type of response from copyright holders, adapting to a new audience and their viewing habits. This seems to be a desirable outcome for both parties, with Youtube retaining the content and its viewers and the copyright owner receiving some proceeds from the performance.

However, if we look at some trends in the music industry, which I believe can provide some insight into the future of film and gaming industries "ad-supported streams are nine times less efficient than paid-for subscriptions in generating income" (Spence, 2015, citing Doug Morris, Sony's CEO). As of June 2015, there are 5 major streaming services that are converging on a very similar offering. These services are Spotify, Rdio, Google Play Music, Tidal and Apple Music, and all of them are offering an unlimited subscription which costs ~ USD 10. At this price point they all offer pretty much the same content.

Another argument supporting the proliferation of streaming services comes from the latest data in internet bandwidth usage. Even if there is little data available outside North America, I think this analysis is still useful because this region has the highest availability for content distribution. The report is by no means exhaustive, but according to Sandvine, is "based on a representative cross-section of Sandvine's data from a selection of Sandvine's 250-plus customers spanning

North America, Europe, Middle East and Africa, Caribbean and Latin America and Asia-Pacific" (Sandvine, 2015). According to them, Netflix traffic is up to "36.5% of downstream traffic in the peak evening hours" while "BitTorrent continues to see a decline […] and now accounts for only 6.3% of total traffic in North America" (Sandvine, 2015). This would suggest that, where available and reasonably priced, streaming services are preferred to free but possibly copyright infringing alternatives.

As of June 2015, Netflix is available in United States, United Kingdom, Canada, Brazil, Australia, New Zealand, France, Germany, Netherlands, Switzerland, Austria, Ireland, Luxembourg, Belgium, Denmark, Finland, Iceland, Norway, Sweden, Brazil, Argentina, Chile, Colombia and Mexico. I think that affordability is an important aspect when dealing with piracy and it is worthy to note that Netflix is present in 17 out of the top 20 countries featuring a high level of household income. Another interesting aspect is that almost a third, 29% of its users are accessing the service via a VPN (Mander, 2015) which enables them to watch Netflix from regions where it is not officially available, or it allows them to access content from other regions. According to Alex Hearn of The Guardian, "more than 30 million Netflix users live in countries where the service is unavailable" (Hern, 2015). The article however, makes a clear distinction between user and subscriber and claims that account sharing is a common practice among these users. This is a grey area because Netflix should not stream content to regions not covered by their licensing agreements, and should not allow subscribers to register from countries where the service is not officially available. However, I think Netflix is simply acknowledging that consumers demand a fair distribution system which does not have regional barriers, a vital aspect of my proposed model.

Another confirmation that is a growing market comes from broadcasters and content distributors who are responding with their own streaming services. Sling TV is such a response coming from Dish Network, a traditional satellite TV provider. An over-the-Internet live-TV streaming which comes as a $20 subscription fee for 12 of the most popular basic-cable channels in the US, including Cartoon Network, CNN, Disney Channel, ESPN and ESPN2, Food Network, TBS, and

TNT. There are also two $5 add-ons: Kids Extra and News & Info Extra.  Consumer reports suggests that this line-up "is the closest service yet to à-la-carte cable television" (Consumer Reports, 2015). Without a doubt the Dish Network leveraged their relation with broadcasters and is offering their traditional, linear content within a new distribution medium. The opposite is true as well, where traditional broadcasters are complementing their over-the-air delivery system with non-linear streaming services, like 9jumpin.com.au, a free website which Channel 9 in Australia is using to stream its content, interrupted by commercials, of course.

Australia is a very interesting case where a combination of neglected audiences and a relatively high availability of broadband services gave rise to the "world's most proliferate pirates of movies and trendy television shows" (Moncrief, 2015). The arrival of Netflix in this space has "resulted in a drop of about 25 per cent in the rate of piracy" (Moncrief, 2015). From November 2014 until July 2015 the percentage of "Australians pirating content at least once a month has fallen from 23 per cent to 17 per cent"  and at the same time "the percentage of people who said they never watch pirated content increased, from 67 per cent to 70 per cent" (Moncrief, 2015). It should be noted that even though two years ago, Australia had a choice of Quickflix, iTunes, Foxtel Play, Bigpond Movies, Google Play, Fetch TV and Mubi, (Jager, 2013) all services capable of streaming content, in June 2015, Netflix already had more than 1 million customers, almost four times the combined subscribers of Presto, Stan, Quickflix and Foxtel Play which come in at 271,000 (Moss, 2015).

However, the most recent and disruptive example is a relatively obscure application called Popcorn Time. It was a free, multi-platform, media player which could source and prioritize torrent content in order to achieve a Netflix-like experience from illegal sources. According to torrentfreak.com, the original software development was shut down by the Motion Picture Association of America, but the project, an open source development, has been picked up by numerous other developers (Van der Sar, 2014). The proof of concept is out there and almost any programmer can create a similar program or build upon the existing code.

The living room has seen its share of technological developments and battles over the years, from television sets, black and white vs colour, gaming consoles, Betacam vs VHS, and now we're witnessing the latest round in this continuing saga. This is a very dynamic scene, and we see the old technologies struggling to remain relevant and failing to meet the expectations set out by a new generation of consumer. We also see new technologies being tested and refined, some of them failing and coming back in a revised form, and some enjoying widespread adoption and success.

### 3.3. What's it gonna look like?

Many factors can influence the design and construction of a house, from climate to building codes and regulations, from community concerns to cost, so the usual approach to such an undertaking is to evolve or repurpose an existing concept/design. However, from time to time, we find experimental designs, radical departures to the traditional concept of building, furnishing and managing a house. The Monsanto House of the Future, Xanadu Houses and the INTEGER Millennium House are just three examples of such experiments.

The Monsanto House of the Future was built as an attraction for Disneyland, and was on display in the California based franchise, from 1957 to 1967. It featured a reinforced polyester structure, fibreglass surfaces and furniture, alongside other innovations, such as variable transparency polymer ceilings and windows, a CCTV system, an ultrasonic dishwasher and a microwave oven. In the 1940s, Monsanto was a major manufacturer of plastics, most notably polystyrene and other synthetic fibres and this concept was built around their capabilities.

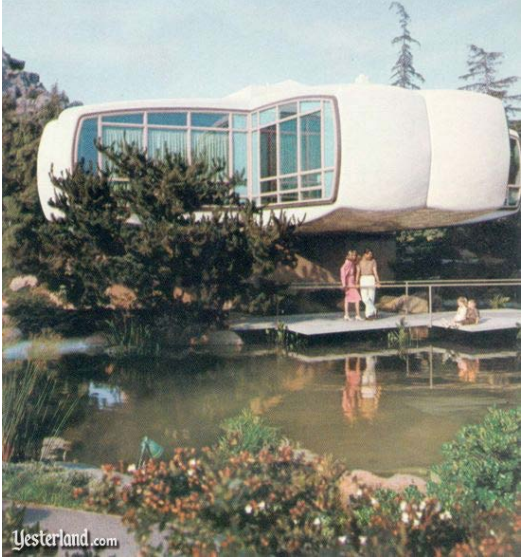

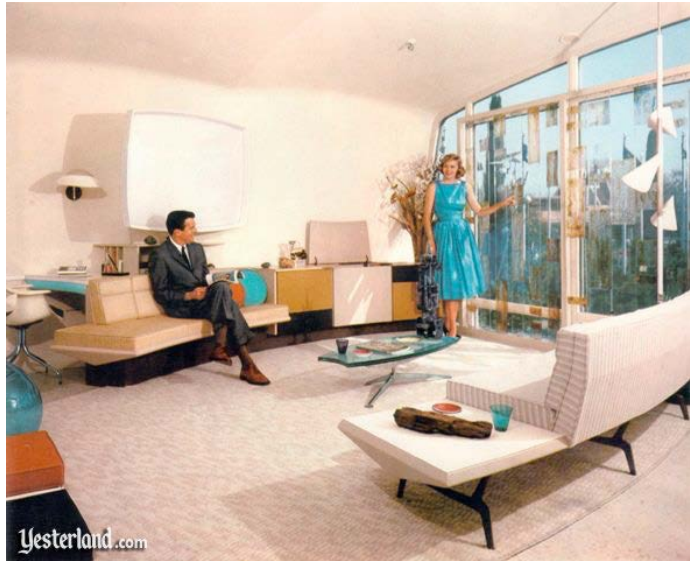

Figure 1. Monsanto House of the Future - exterior (Yesterland.com, 2015)

Figure 2. Monsanto House of the Future - living room (Yesterland.com, 2015)

New living room furnishings match flowing curves of house, feature upholstery and carpet of "Acrilon" urethane foam cushioning. Powered, revolving louvers of plastic screen beside window cast light patterns on TV-movie-stereo center along wall." (Monsanto, 1960)

Another example are the Xanadu houses. These are a series of concepts built in the early 1980s, featuring a structure of polyurethane insulation foam which would require very little time to build. They also featured an early example of home automation, even home intelligence:

Take Xanadu's kitchen, for example. It's equipped with a "family dietitian" consisting of four microcomputers. It plans well-balanced meals for family members depending on their height, weight, sex, age, and levels of activity. If you come home from a busy day and inform the computer-dietitian that you skipped lunch and nibbled on a candy bar instead, it calculates supper based on the nutrients you missed. An "auto-chef" can move food from the refrigerator to the microwave oven to the dining table, and the computers keep track of the grocery inventory so you know what to replace. The auto-chef can even regulate the ambience of the dining room to match your meals, adjusting the lighting and background music to complement your Mexican dinner, for instance. (Halfhill, 1982)

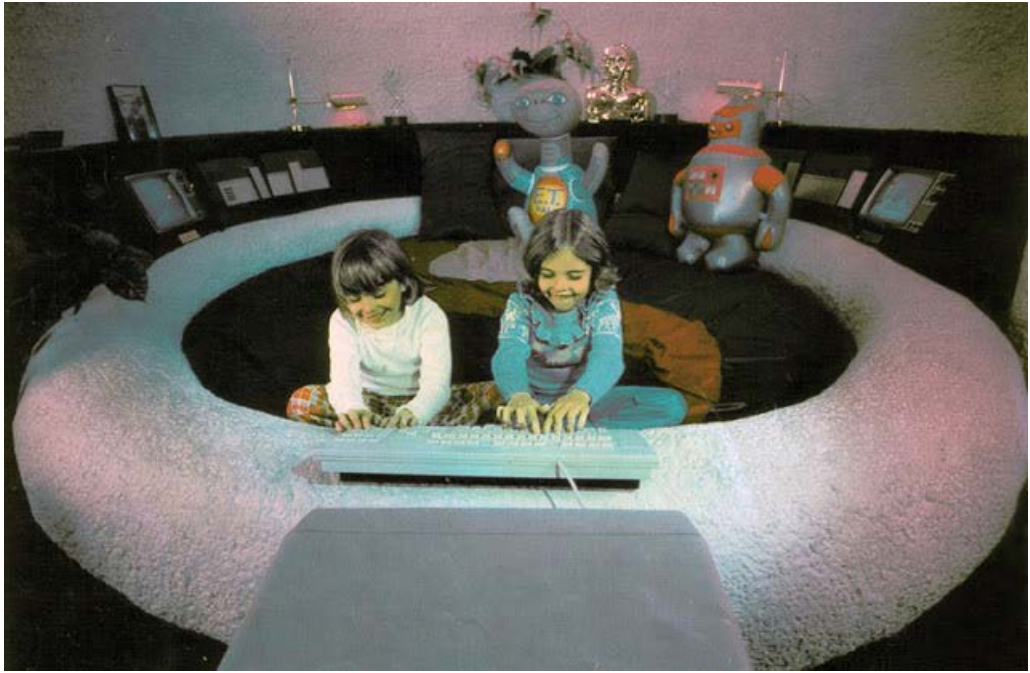

Figure 3. Xanadu Homes - bedroom (moquetadesign.blogspot.com.au, 2015)

In an article, which appeared in the December 1982 edition of Compute!, Mason Roy, the architect of the Xanadu home build in Kissimmee, Florida, also describes the "electronic hearth": "a home computer that is the center of the family's activities - entertainment, bookkeeping, meal-planning." (Mason cited in Halfhill, 1982). The list of proposed concepts is quite extensive: video displays featuring computer-graphics art, a home office with a computer able to access electronic mail and news services, teaching microcomputers, videotextured windows, an environmentally-controlled habitat with fire and security systems. But, as the article faithfully predicted, "the biggest hurdle [was] market resistance from people unaccustomed to delegating tasks to computers" and the Xanadu houses grew less and less relevant as the technology aged. The last one was closed to the public in 1996.

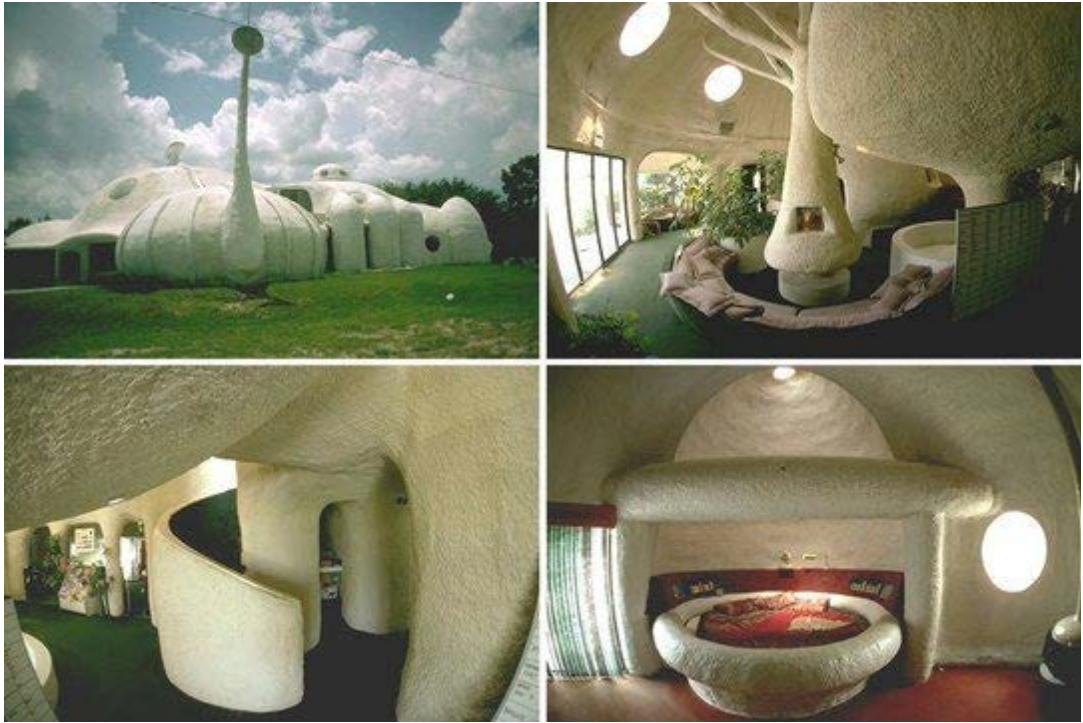

Figure 4. Xanadu Homes  - various views (Hugh, 2012)

The INTEGER Millennium House is a demonstration house, located in Watford, England. Opened to the public in 1998, it was set up as a demonstration project without a set budget, open to suppliers donating expertise and materials. The INTEGER Millennium House was built using only standard components, most of them prefabricated. It featured a green roof, a wind turbine, solar photovoltaic panels, solar water heaters, a geothermal heat pump, rainwater collection and a grey water recycling system. The home automation technologies include a building management system in charge of heating, the automatic garden reticulation, a security system which uses microchip-embedded programmable door keys and 4 predefined lighting modes. It received significant media coverage, most notably through the "Dreamhouse" series on BBC1, and claims more than 5000 visitors, many of which "wanted to know where they can buy an INTEGER house, and why housebuilders are not offering this kind of product to the public" (Building Research Establishment, 2004).

Unlike the previous examples, the INTEGER house was given a second lease on life and refurbished in 2013. This is one of the most interesting aspects of this project, because it provides expertise in retrofitting new, energy efficient technologies and materials into an already existing space. The retrofitting vastly improved the efficiency of the house, with a new Air-Source-Heat-Pump, Building Integrated Photovoltaics, Solar-thermal Honeycomb Collector, a roof light which can be programmed to release trapped heat and led lighting, all controlled and integrated by a home automation system.

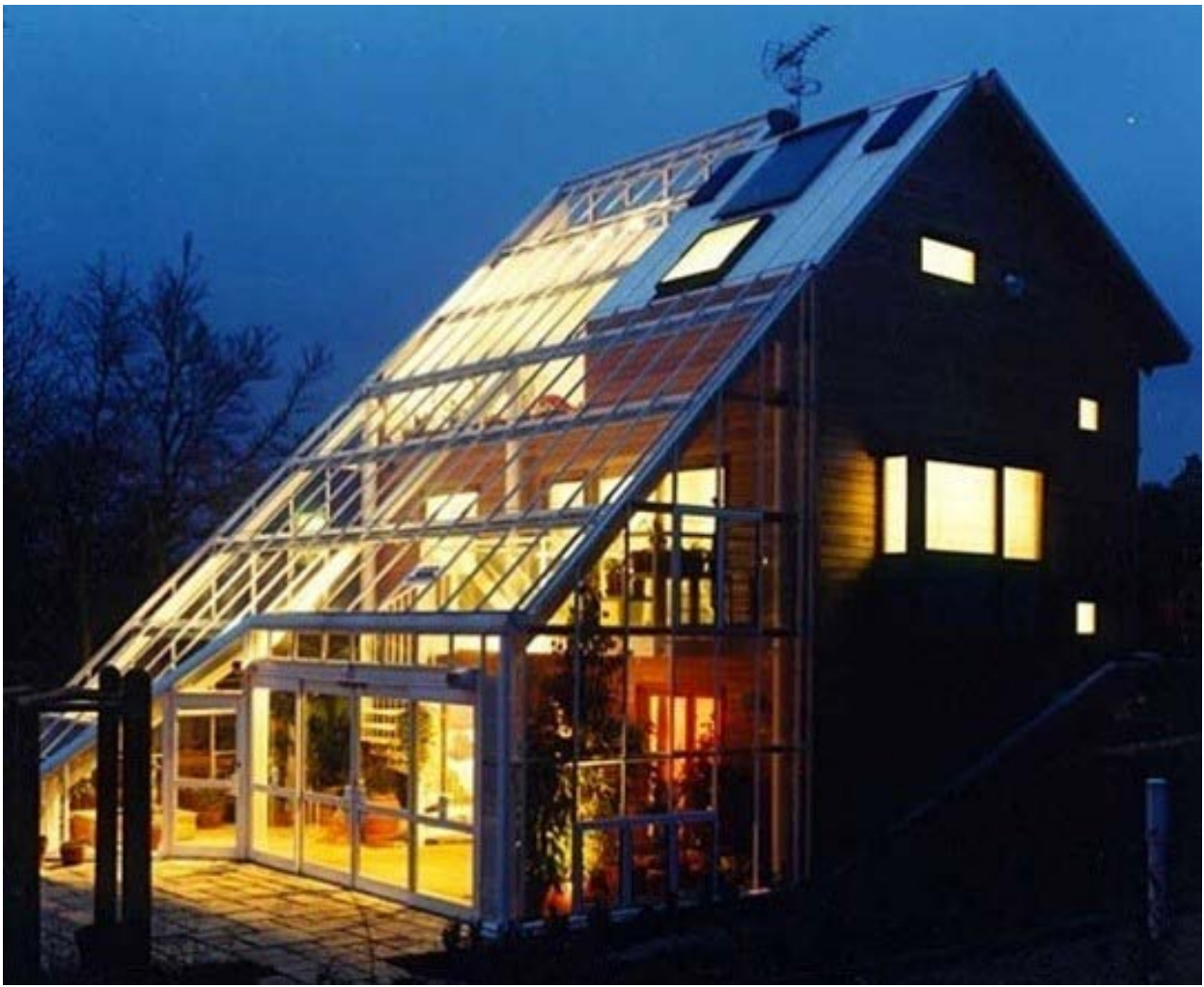

Figure 5. INTEGER house  - exterior / full lighting (Thompson, 2008)

The home automation system comprises intelligent, app-enabled heating controls and lighting sensors that have daylight and motion detection to regulate lighting levels. It uses the global KNX standard so different appliances, devices and systems can communicate with each other (Building Research Establishment Ltd, 2015).

I think that something has changed since the 1950s, when some of these smart homes have been imagined. A crucial difference is our attitude towards computers. We have incorporated them in almost every device we are using today: cars, thermostats, phones, Television sets, etc. Also, we are finally reaching a certain comfort level with technology, and we are not afraid to buy and use it. Another significant difference is the level of interconnectivity among the various devices. The Xanadu homes required extensive cabling installed in the building phase. In comparison, the managing system used in the INTEGER house, a Siemens QAX 913 uses a KNX RF-compatible, 868.3 MHz bidirectional wireless connection as well as a more traditional, wired, KNX TP1 protocol. It can be controlled and programmed via a PC or iOS and Android apps.

Finally, there is a very compelling argument made by the level of efficiency achieved using home automation, which ultimately translates into significant cost savings. The Xanadu homes were criticized for their power-hungry computers running 24 hours. The Siemens QAX 913, is rated at 7VA, which should cost (based on a UK rate of 20c/KWh) about $5 to run continuously for a year. Traditionally, automatic control of the environment was developed for large buildings, where the initial cost could be easily offset by the scale of the efficiencies achieved. I think this cost is finally within reach of single-family dwellings, in part because the traditional vendors are seeing a tremendous pressure from open source projects like Arduino, a low cost, hardware and software solution that can be easily implemented in almost any development. Another incentive to increase our energy efficiency comes from the demand for electricity in the future, driven by the anticipated adoption of electrical automobiles. There is a need for a well managed and efficient system, capable of charging the car's batteries, using them to supplement additional energy demands, or even acting as backup generator.

The INTEGER house serves as a beautiful and fully functional example of what can be achieved today in all areas defined by Björkskog (2007): welfare, environment, entertainment and communications. However, this house is just an expression of the builders capabilities and we do not know how it would address a consumer demand.

According to Transparency Market Research, "global home automation market was valued at USD 4.41 billion in 2014, growing at a CAGR of 26.3% from 2014 to 2020". A reverse calculation based on a 26.3% CAGR (Compound Annual Growth Rate) yields a market size of 17.860 billion in 2020 (Transparency Market Research, 2015). The report also mentions that "safety and security segment held the highest market revenue share in 2014", and that "the home automation market is segmented on the basis of application into lighting, safety and security, HVAC (Heating, Ventilation, and Air-Conditioning), entertainment, energy management, among others"; a further segmentation is done by market: (1) luxury home automation systems, (2) mainstream home automation systems, (3) DIY (Do It Yourself) home automation systems and (4) managed home automation services (Transparency Market Research, 2015).

The luxury or custom market installation is usually achieved by contracting the services of a system integrator. These companies match their client's needs to the current capabilities of various manufacturers and implement, program, test and service the system. An example of such system is Ian Poulter's Florida residence. Tracy Adcock of Orlando-based HSS Custom AV integrated various products from Crestron, Axis, Lutron, Sunbrite, Vutec, and others to deliver a custom home automation system for his 6000 square foot home (Montgomery, 2013). The installation features 16 Crestron touch panels around a Crestron control system. Each panel along with an iPad app, controls temperature, lighting and automated sunshades, functions as intercom and baby monitor, provides video feeds from the 6 bay garage and assists in opening the appropriate garage door. A Crestron DigitalMedia HD video distribution hub stores the content and feeds it to any screen on the premises whether outside or in the kids bedroom. A 12-seat home theatre with a 123-inch screen features three gaming systems, PlayStation, Xbox, and Wii, a Blu-ray disc player, DirecTV receiver, cable TV receivers and an Apple TV.

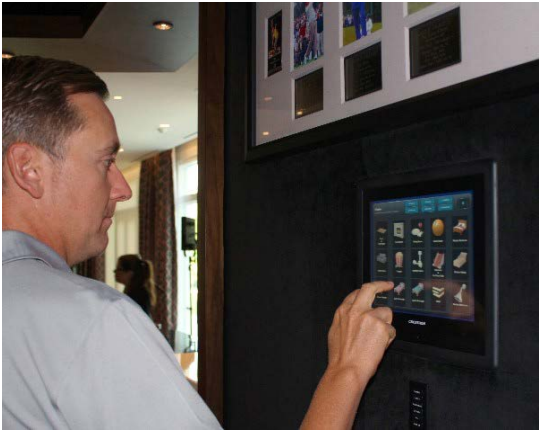
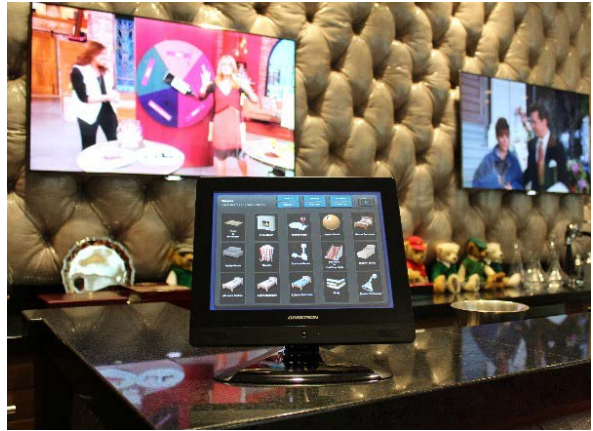

Figure 6. Panel can be used as intercom. (Montgomery, 2013)

Figure 7. Crestron Touchpanel (Montgomery, 2013)

A Crestron Sonnex multiroom audio distribution system is controlling the music throughout the residence. Everything can be controlled and customized locally or via internet, and includes preset mood scenes, such as a party mode when "LED lighting accentuates trophy cases and illuminates a stunning glass and stainless steel staircase while activating the music system" (Montgomery, 2013). This is a fairly typical installation tailored to the very specific needs of a client.

An alternative approach is Zac Adams' apartment, an example of a mainstream system, which bypasses the knowledge and expertise of a system integrator and offloads these tasks to the consumer. Zac's 1,180 square foot living space, features 6 android tablets acting as touch panels, "follow me" music throughout, temperature control, automated lighting, built-in security system and a scent controlling system.

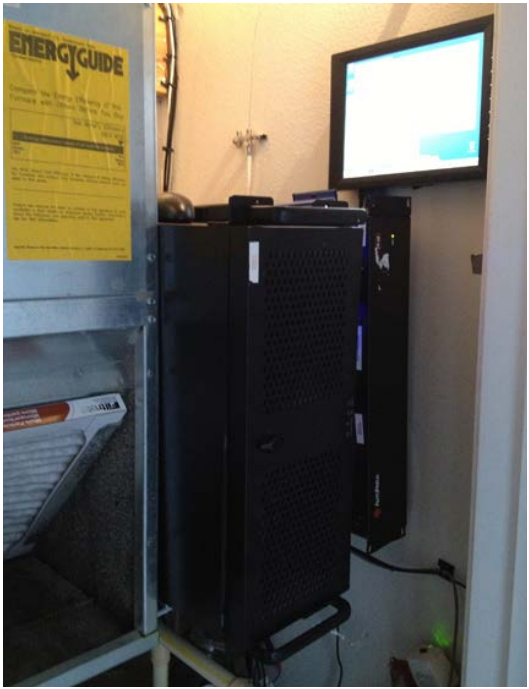

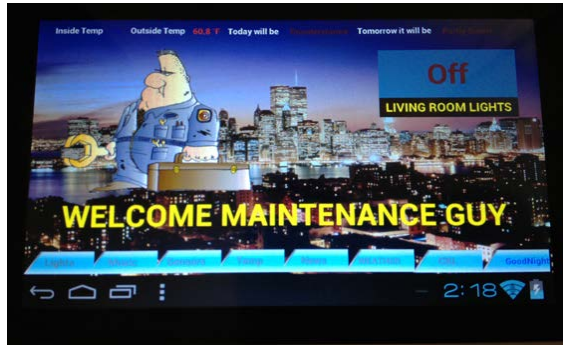

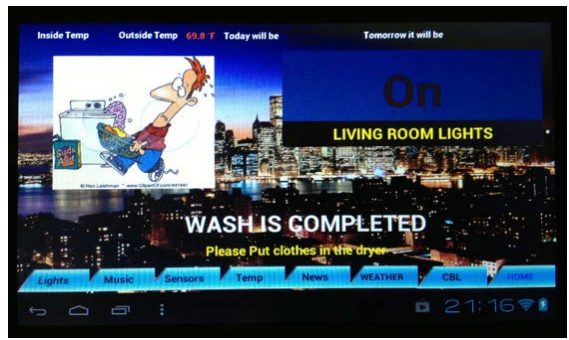

Figure 8. Zac's utility room with HomeSeer server (Cericola, 2013)

Figure 9. Different alerts on Android tablets (Cericola, 2013)

Zac is using HomeSeer's software to control a "mixture of Z-Wave, Bluetooth, Wi-Fi and X10 devices" (Cericola, 2013). The entrance is controlled by a Yale Real Living lock, controlled by key, touchpad or his iPhone. The HomeSeer server also allows Zac to implement his own programming, via scripts and monitor news and weather with audio and visual alerts. A notable feature is a 5 minute timer for his closet doors which prompts the system to send out a reminder every minute until it's closed. "It's a great training tool for kids […] It's very good if you have a family, because HomeSeer becomes the bad guy — and computers do not give up" (Adams Z., cited in Cericola, 2013). The system also features automated "fountains and lighting on the patio, a video camera that can track motion and send pictures to Zac's iPhone, and an "Away" mode for the lighting, temperature and AV equipment" (Cericola, 2013). Zac also unloaded some of his everyday tasks to the HomeSeer server:

… if the system detects motion after 6:30 a.m., it will trigger the coffee maker. It's also connected to chore-related items, such as a washing machine and dishwasher. The control system can measure power being used by any of these items and when the cycle is complete, it sends a text message and/or makes an announcement over the apartment's speakers. (Cericola, 2013)

This system, in its current form, took about 18 months to build. It might seem a long time, but it's perfectly justifiable by the amount of research, troubleshooting, trial and error Zac had to do on his own.

The DIY market segment, refers to mainly user developed hardware and software solutions and requires the most expertise and commitment from its builders/users. It is driven mostly by hobbyists and affordable open source hardware like Arduino or Raspberry Pi. When coupled with open source software, such as Home Automation Server or Open Source Automation Server these platforms become a viable option to "off the shelf" systems. The user base is constantly innovating and sharing: Wireless Internet Thermostats, Sonic eyes (allow visually-impared people to walk without a cane), Internet controlled cars, Segway-like balancing robots, remote controlled desktop missile launcher and even an automatic control system for a tiny Ni/H fusion reactor (Arduino, 2015). All the hardware required to build these projects, step by step instructions, and the software, are readily available at very affordable prices. More importantly, DIY home automation provides a benchmark for both consumers and manufacturers in terms of design, price and availability.

From a consumer point of view, the fourth segment, managed home automation services, might not be different than mainstream systems except they would pay for it as a service (monthly fee) and not upfront as a product. These are extended offerings from established Telco operators or cable providers, usually offered in bundles and feature capabilities like: Remote Access, Instant Alerts, 24/7 Central Monitoring, Wireless Two-Way voice, Touchpad controls, Lighting & Small

Appliance Control, Thermostat Control, Live Video Streaming, Smoke/Fire Monitoring, Carbon Monoxide Monitoring, Water Leak Monitoring (Rogers Communications, 2015).

In a nutshell, this is what home automation looks like today, a fairly sophisticated system built around two simple concepts: remote control and task based programming. However, these are simply technological relics from industrial applications and there are other examples more innovative and relevant for this project.

In Future Home Design: An Emotional Communication Channel Approach to Smart Space, (2013), the authors go beyond traditional use of interfaces in a smart home setting and explore the concept of using "every-day objects to create communication channels between spaces and people, which can then strengthen interpersonal relationships" (Huang et al., 2013). They implemented a smart space, the Time Home Pub, which included a whiskey glass, an interactive table, a digital picture frame (called LiveFrame), an MP3 player and 2 video projectors. When it was showcased at the Taipei Fine Arts Museum in 2007, the system was controlled by three different computers and programmed to respond to a specific scenario:

| | Trigger | Response | Result |
|---|---|---|---|
| 1 | Sean places a whiskey glass on the interactive table | The system to change into the bar mode. "The environmental lighting slowly becomes darker, and the wallpaper switches to an animated pattern" and "simultaneously, time-marks and emerge on the table to stands for Sean's friends and family who visited the space in the past two years" (Huang et al., 2013) | Sean remembers "friends and family who visited the space in the past few years", including Sasada and "realizes that they have been apart for a long time" (Huang et al., 2013) |
| 2 | Sean moves the whiskey glass on Sasada's time-mark (a friend he did not see in a long time) | The system loads photos with Sasada in the LiveFrame and starts to play a slideshow | The photos trigger Shawn's personal memories with Sasada |
| 3 | Sean moves the whiskey glass on the music hot spot | "music gradually begins to play" and "the pattern flow on the interactive table follow the musical melody" (Huang et al., 2013) | The environment is adjusted to better suit Sean's preferences |

Huang et al., exhibited this concept for 2 months in 2007, and concluded that "most visitors were interested in this future home design" (Huang et al., 2013). The visitors raised questions about the system's flexibility and feasibility, inquired about additional modes and the possibility of linking two or more smart spaces in real time. The authors also remarked that "compared to the visitors who did not read the instructions, visitors who did, easily adapted to the scenario and asked further questions about feasibility" (Huang et al., 2013). Their concluding remarks suggest that "smart spaces not only provide a more natural communication between users and space, but their design must also take user-friendliness into consideration" (Huang et al., 2013).

In a more recent study, Yu-Chun Huang and Scottie Chih-Chieh Huang (2014) present an evolution of the "Time Home Pub" concept, called the "Personalized Smart Living Room" where multiple users can interact in the same time with each other and with this space, a concept which provided some impetus for my own design. A new feature is the use of smartphones and their integration in the smart space. They describe three different scenarios, two individual interactions and one where two people are present.

| | Scenario | Response | Result |
|---|---|---|---|
| 1 | Saori comes home, sits on the smart sofa and starts browsing through photos on her smartphone | "The sensors of the sofa recognize that Saori is now tired, the room's interactive wallpaper displays Pop Art, and the room plays Saori's favorite music from her smartphone to help her relax. Since Saori is browsing through photos she has taken with her cell phone, the table transforms into a large photo frame and displays a slide show of photos to provide more comfortable feedback" (Huan, Y. & Huang S., 2014) | The system is personalized to Saori on both a visual and audible level. The system is also augmenting the user experience by facilitating the use of a larger display. |
| 2 | Scottie comes home and sits on the smart sofa. The interactive table alerts him of a change in his schedule. | "The wallpaper responds to his posture by presenting an animated pattern on the wall. Meanwhile, the room plays Scottie's favorite music from his smartphone. Suddenly, he notices that there is an important unread message on the table" (Huan, Y. & Huang S., 2014) | The system is personalized to Scottie, again, on a visual and audible level. The system is also acting as an extension of Scottie's smartphone and facilitates the use of the interactive table as a display. |

| | Scenario | Response | Result |
|---|----------|----------|--------|
| 3 | Saori and Scottie come back home together and sit on the smart sofa. | "In order to reflect the joyful and harmonic atmosphere, the environment changes to its bar mode. Spots on the wallpaper change color according to the users' movements, and the space plays Jazz music" (Huan, Y. & Huang S., 2014) | The system is able to recognize a social situation and acts accordingly, adopting a neutral setting, which presumably is preferred by both users. |

The concept was not featured publicly and thus there is no feedback from the public. What Huang et al. set out to do was to demonstrate a new inter-relationship framework in a smart space, where ubiquitous computing is augmenting the physical space in form and function and where HCI (Human-Computer Interaction) is taking place through the user interaction with his or her surroundings, not within a pre-programmed interface. They conclude that:

> In the 21st century, architectural design must take into account "ubiquitous computing". Architects have to not only consider the exterior forms, but also a building's interior functions. They must also keep in mind how to adequately merge technology into our lives. Therefore, HCI in a smart space should primarily be implemented based on its architectural context and human needs. (Huang et al., 2014).

I think the most significant aspect of the "Personalized Smart Living Room" is the integration of smartphones to facilitate some of the customization features. With more and more features added continuously to these devices, such as health monitoring, activity tracking, payment methods, GPS tracking, etc., they also become a central storage for our personal data, our preferences, our habits. It makes sense to use that data and make the interaction with our surroundings more organic and fluent. The examples shown, streaming music and calendar notifications are just the tip of the iceberg when it comes to using a smartphone as an HCI link. Our contacts, family, friends and business partners, the frequency of calls and messages, our daily routines, from eating habits to transportation can be shared and used to "prime" our interactions within a smart space. Of course, this type of data sharing would raise a few privacy concerns, which Huang et al.'s prototype did not address in any way.

All the commercially available products mentioned before, in all four market segments, featured extensive use of smartphones and tablets. However, I think the market predictions made by Transparency Market Research (17.860 billion market size in 2020) were based on current capabilities and cannot account for the introductions of new technologies, such as wearable devices, which appeared around 2013. It took around three years for tablets and smartphones to mature and be used in home automation systems. As a matter of fact, it wasn't until 2014, when Apple, a major manufacturer of smartphones and tablets introduced HomeKit, "a platform that allows devices in the home to 'talk' with Apple products" (AppleHomeKit.com, 2015). HomeKit came along with iOS 8 and introduced, arguably its most important feature, the capability of defining time based or location based custom triggers.

Wearable technologies, in their current form, glasses, smartwatches and activity trackers, are relatively new and still in their infancy. Despite a very optimistic Google Inc. and a very enthusiastic following, three years later, only 1% of Americans have spend $1500 on Google Glass (Adweek, 2015). Activity trackers were found with 11% of Americans, and smartwatches, the newest category in this market with only 3% (Adweek, 2015). However, there is a large discrepancy between these wearables when we look at the user base split by gender: activity trackers are split fairly even, with 54% women and 46% men, while Smartwatches are split at 29% women and 71% men (Adweek, 2015). This suggests that, in 2015, smartwatches are still situated early on the adoption curve, with more potential to grow than activity trackers. But ultimately, critical to the success of wearables will be the fact that 50% of the U.S. population does not wear a watch at all (Hold E., cited in Adweek, 2015). This would imply that smartwatches need to deliver new capabilities and features to see widespread adoption.

I think smartphones triggered a technological revolution which is redefining our relationship with technology and our surroundings, including living rooms. There are other forces at work in this space, an increasing pressure to use less energy, new materials and new construction techniques. But ultimately, we are increasingly depending on computers for our security and physiological needs, the very bottom of Abraham Maslow's pyramid. In 1943, Maslow proposed a theory of human developmental psychology, where he claimed that our minds and brains are constantly trying to balance five different types of needs and subsequently, the motivations behind them. These are our most basic and immediate needs, such as air, water and food, followed closely by safety, of all kinds, physical, economical, etc. Then we have love and belonging, followed by esteem and self-actualization (Maslow, 1943). I believe that only recently, with the advent of smartphones and wearables, we have started to incorporate, in a conscientious and deliberate manner, computer technologies in our most basic needs: air, water, food and safety. And because these needs are very important to our psyche and critical to our survival, the issues surrounding this adoption are higher than ever, certainly worthy of a review for existing practices if not a different implementation all-together. We are not talking about kids playing computer games anymore, we are literally becoming dependant, as individuals and a species, on computing technologies, which raises very important issues in terms of identity and human rights. Furthermore, I think that we are losing our privacy and individuality as a direct consequence of the way in which these technologies are implemented.

I think our living-rooms, this public and private space, weaves all these new technologies into our family lives, bringing our digital worlds much closer than ever before, and that is an experience that we are going through every day.

### 4. A new model

Question #3 - What alternative and/or augmented model for the future living room and screen entertainments could be proposed - what legal and economic barriers might resist this and how could they be overcome in an alternative model?

The model presented in this chapter is designed to support rather than test the ideas put forth. As such, the consideration behind the design is the experience itself and not the feasibility of this environment. This model can be built with current technologies, but some of them or even the concept itself, may not be considered a priority or even economically feasible by those who might have such capabilities. Nonetheless, I hope it can serve as an example to improve the experience, the efficiency and to give consumers of content and technology a fair choice.

### 4.1 Architectural considerations.

A quick revisit of Rechavi's list of described living room activities, such as entertaining, watching TV, listening to music, talking on the phone, working, studying, reading, eating, meditation, relaxation, contemplation, conversation, exercising, having sex, napping and occasional sleepovers, reveal the versatility which we expect from this space. Other than preparing meals, which requires specialized equipment, virtually everything else can be happening in the living room. This is, without a doubt, the most challenging aspect of this project, creating a space not only capable to customization, but one that invites it and transforms with it. In addition, the living room should also support its users emotionally, individually and as a family.

The space incorporates two distinct areas: the lounge and the office. This is a configuration which allows all the activities mentioned to happen in the same space and sometimes it enables two or more to happen in the same time. Both areas feature extensive shelving to support personalization and the display of meaningful objects, and walls support both physical and digital objects.

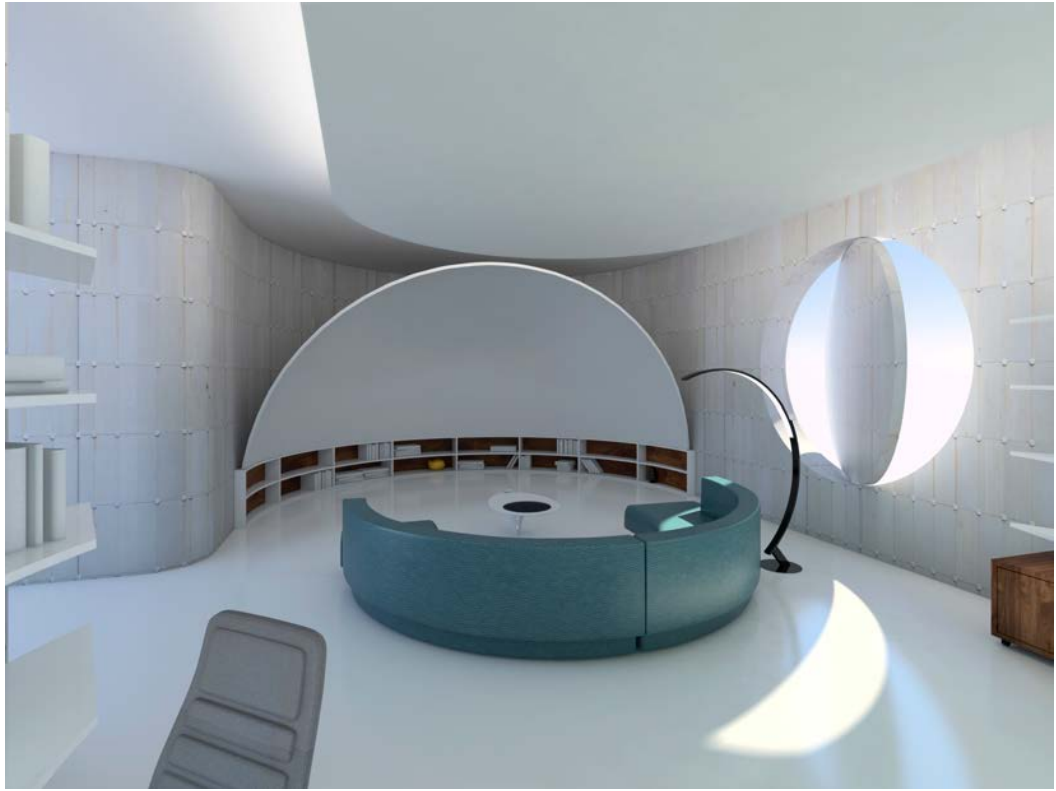

Figure 11. Lounge side of proposed model.

There are several distinct architectural and design features integral to this concept. First is an automated control for natural sources of light. The purpose is twofold: controlled lighting can improve the image quality rendered on the various screens present and secondly, and allows for a fully customizable lighting environment (intensity and colour temperature). This is achieved by the second feature, the false ceiling. This is a simple and effective way to "hide" much of the technology present in this room while still having easy access for service, upgrades, etc. The false ceiling would provide the perfect housing for speakers and video projectors along with the primary lighting sources. On a day to day basis the system would simply provide either neutral, consistent, context appropriate illumination, or enhance the mood as desired, from mimicking a bright sunny summer day to a campfire. The integration of lighting control within other activities, such as watching television or playing a video game can also play an active role rather than a passive one. If you are watching a concert, the system can buffer and delay the program a few seconds to analyze and mimic the concert's lighting in the room. Of course it can also serve as a warning mechanism in emergency situations or to assist people with disabilities.

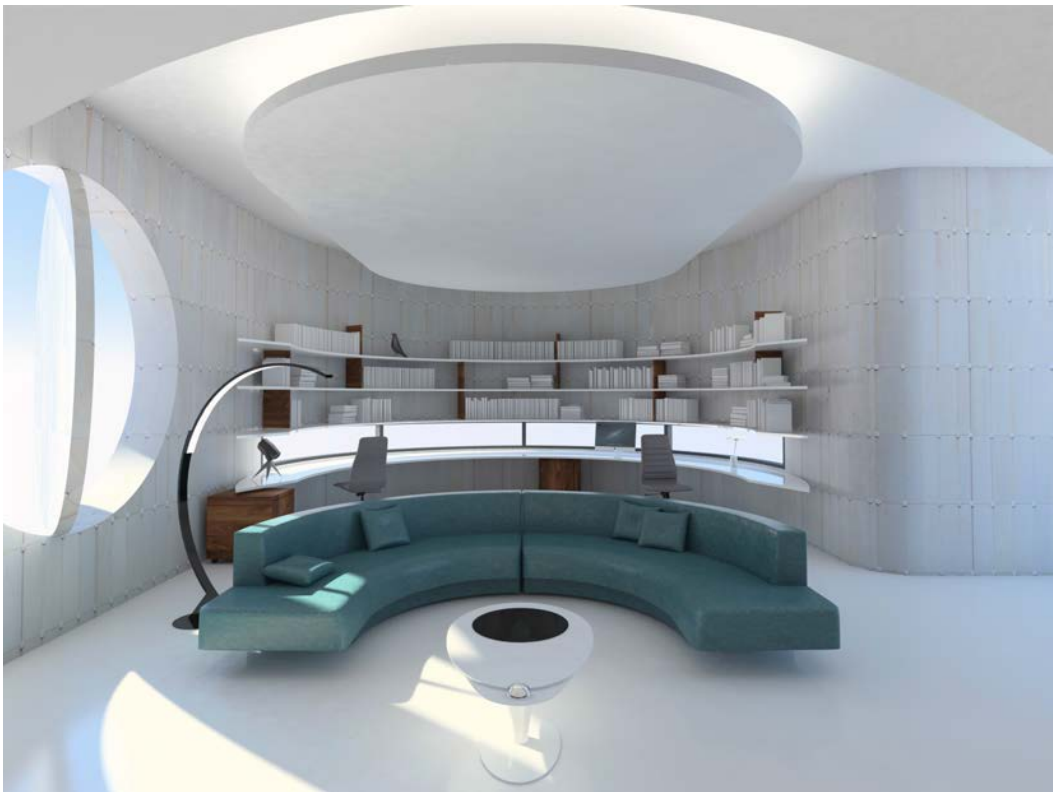

Figure 10. Office side of proposed model.

The third feature is the large dome shaped surface present in the lounge section. Its primary purpose is to be used as a projection surface as described in Paul Bourke's Mirrordome. He proposed "an alternative full dome digital projection system […] based upon a single projector and a spherical mirror to scatter the light onto the dome surface (Bourke, 2005). It offers many advantages over the conventional systems, and provides a similar quality of projection at a very reasonable cost. I think the most remarkable feature of Bourke's concept is how easyly it can be standardized, using different mirrors to achieve an ideal projection coverage and increasing its suitability for various room sizes.

Other distinguishable features are the smart coffee table and the couch. The coffee table doubles as a display, control surface and houses the mirror used for the dome projection. The couch is a modular, mobile design which allows for multiple positions and configurations, whether self-powered and controlled by a computer or simply able to be repositioned easy and securely.

Overall, the space should be user friendly and familiar, and needs to retain all the traditional controls, such as light switches, window opening mechanisms, etc. in place, in parallel with the automated and remote controls. I think this is a critical aspect of implementing new technologies in the living room, especially those which seek to replace or mediate human commands.

### 4.2. Software considerations.

I believe this is the main reason holding back technological development in many areas, including home automation and entertainment systems. We have seen examples of "smart" furniture or appliances, such as the ones described by Park et al. in 2003: smart pen, wardrobe, dressing table, bed, pillow, mat refrigerator, etc. Some devices are enhanced with new, complementary functions, like a smart mat gathering information about the person waiting by the door, weight and footprint which is used to identify that person,. Others are improved versions in a more convenient or practical design such as remotes "capable of recognizing an object via an internal camera,  promptly displaying an appropriate GUI for any system in the smart home" (Park et al., 2003). Some of these concepts were even brought to market, like LG's Internet Digital DIOS refrigerator, announced in June 2000. Telecompaper, a well respected research and publishing company claims that LG "has invested [South Korean] Won15 bil into R&D of the Internet fridge since 1997", roughly $50 million US dollars (Telecompaper, 2000). And here we are, more than a decade later, and the Internet fridge, like the rest of those appliances and devices are nowhere to be found in our homes. I believe the reason why many of these devices have remain just concepts is the absence of an environment which can support and interface with them.

Logitech's latest Harmony remote, the Elite, promises to "streamline life with intuitive, integrated control of both home entertainment and home control devices"(Logitech, 2015). The remote itself is now a "universal remote, hub and app", and works like this: "the included Harmony Hub sends commands from the remote or the app using IR, Wi-Fi® or Bluetooth wireless signals" (Logitech, 2015). It promises to control over 270,000 devices, from more than 6000 manufacturers. It needs to be set up, customized with your devices and activities based on templates suggested by Logitech. Sadly, this universal remote, which does not control any devices but rather a specialized hub, is the closest we have come yet to developing a common platform for all devices and appliances in our homes.

I think the underlying philosophy of such system, its basic set of rules still has to be defined and agreed before any significant advancements will be made. I think the system has to fulfill 3 basic rules:

1. The system can be bypassed by conventional controls such as light switches, locks, etc. Furthermore, the points of failure should remain unchanged. For example, if the light switch does not turn the light on, there can only be three points of failure, a) there is no power, b) the light switch is faulty, c) the bulb is burned out.

2. The system is capable of learning and adapting to its users without additional programming. In other words, the system is not dependant on preprogrammed step by step instructions. Only the learning capabilities are programmed into the system and all the subsequent data is acquired from usage and behavioural patterns.

3. By design, the system cannot be interfaced with or its data accessed from outside of the premises. This can be achieved either by using protocols that are not compatible with TCP/IP (the current protocol in use) or even using non-binary, non-silicon based computing technologies. However, this does not mean that the system would not have access to outside information, but rather that its own data cannot be transmitted or understood outside the premises.

I think these rules would lay the foundation for a system that can operate safely and efficiently in our homes without compromising our privacy and security. Assuring its customers of confidentiality is the single biggest issue that any company entering this space would face. I have already outlined in a previous chapter, War in Our Living Rooms, a few recent major data security breaches. This is why companies are hard at work, trying to figure out how to solve this issue. Joerg Hartmann, VP of Global Client Computing Devices within the server platform business at Fujitsu outlined some of the innovations they are trying to implement around this issue:

> I think there is a big danger of collecting all the possible data because the question does arise of what the hell we're going to do with it all […] What you'll find a lot more conversation about right now is about a hub that has a bit of intelligence so that only relevant data is routed into the big data centre, where the other stuff may just stay at hub level - basic decisions like heat on/heat off, power on/power off, can be done without cluttering up the cloud" (Hartmann, 2015)

This is what Fujitsu calls "Smart Data", a concept that defines Fujitsu's "holistic approach to the Internet of Things (IoT) revolution" (Merriman, 2015). Hartmann continues:

[… ]we are expanding our data centres, but we're finding it's not so much just because of the amount of data being created, but because we're finding more and more that our customers are wanting to store their cloud data with a Level 3 encryption layer - that stuff we used to call 'hosting' but is now called 'cloud'. (Hartmann quoted in Merriman,2015).

For some companies, this is the most logical approach: encrypt the data, then "hide it" in their cloud along with other data, most likely from other companies or people and hope that nobody other than the owners figure out what is actually stored in there.

I am proposing an alternative to this approach, based on two simple ideas. I think that wearable technologies along with the introduction of an alternative to the common, task based programming structure of home automation, can change the way we interact with our homes and drive a faster adoption of both technologies.

The task-based model has been employed in home automation as a means of defining an action or set of actions to reach a desired outcome. For example, in an office setting, the HVAC and lighting systems can be easily programmed to match working hours. More sophisticated systems also take into account hours of daylight and temperature according to seasons and adjust accordingly. But sometimes, people might need to stay a bit late or come in early, and in this case a re-programming would not necessarily be effective, so the system is also programmed to accept input from motion sensors and manual switches. A similar concept framed in the context of the living room, might look like this: let's define a task as "watch a DVD movie". The task will be programmed on a controlling device as a more complex set of instructions: turn on TV, switch to the correct input, turn on home theatre, adjust sound level, dim lights. Once set up, the system should work reliably until there's a failure or an upgrade, in which case it needs to be reprogrammed.

Both systems however, rely on the skill of their designers and programmers to anticipate and account for as many scenarios as possible. This is a problem we are trying to mitigate as we increasingly rely on automation: "It is impossible to anticipate all the situations a robot might encounter, […] the goal is to advance beyond preset rules to true cognition, and create a robot that can reason through tough problems and justify its actions." (Scheutz cited in Borchers, 2014). A simplistic example in a very narrow application is a GPS. Without a rerouting function, it will fail at the very first wrong turn. The reroute function is built in as an acknowledgement to the infinite number of choices its user has. The programming of such device becomes less task supporting (turn left, turn right, etc) and more outcome oriented.

Recently Apple announced a new generation of their Apple TV which, as far as I know, makes the first transition towards an outcome oriented programming environment.

> "When introducing its TV service, Apple focused extensively on the presence of Siri, its voice-activated artificial intelligence. There's a Siri button on the Apple TV remote. It lets users find content from Apple's library and search within shows, asking Siri, for instance, to rewind a few seconds or to find a particular actor's cameo" (Bergen, 2015)

What makes this feature truly special is Siri's capability to search across multiple catalogs from content providers. We have seen voice search capabilities in other devices like the Amazon Fire TV, but that search is limited to Amazon's content. Siri's voice recognition capabilities are nowhere near perfect but help is on the way. I mentioned earlier the time and location based triggers that one would need to program in order for Apple's HomeKit to control various devices. iOS 9, scheduled to arrive in fall of 2015, will feature new capabilities and support Apple's first wearable, Watch. In this release of its mobile operating system, Apple is rumoured to give HomeKit the capability to recognize user activities and to respond with "trigger scenes (activation or de-activation of a number of HomeKit accessories)" (Etherington, 2015). I believe this is the first evidence, in a mainstream home automation application, of outcome based programming. It is hard to anticipate how successful the new HomeKit implementation would be, but it is a step in the right direction. And HomeKit is apparently working very nicely with Apple's voice command system, or virtual assistant as they call Siri. This is the example provided by AppleHomeKit.com:

The nice thing is that you do not have to give specific assignments, but can also indicate what you will do, for example: 'I'm going to bed. " Siri makes sure that the light goes off, the doors are locked and the heating is put down. (AppleHomeKit.com, 2015)

Wearables should allow these systems to recognize individuals and provide a convenient way to authorize data access and certain tasks. And they can also play a key role in protecting our privacy. We can take the third rule of my model even further, (the system cannot be interfaced with or its data accessed from outside of the premises), completely take away the system data storage capabilities and move the statistical and historical data into our wearables. Not only would we know at any given moment where that data is, but presumably we would also know if that data is being accessed and by whom. And that is because wearables can be anything we want. Right now, it's smart watches and activity trackers, but tomorrow it could be tattoos, wedding rings, any kind of implant, even some kind of biological device that we can carry and think of as a pet… Intel has demonstrated earlier this year a way to use someone's skin to "store information and pass the data from one device to another without the need to send it over WiFi, email or Dropbox" (Bell, 2015).  This is perhaps where the true potential of wearables lies, in a secure, easy to use databank which can customize mostly everything around us, from workstations, to cars, to home environments and entertainment platforms.

The system proposed here relies on wearable devices to provide specific operating modes, depending on factors such as day, time, scheduled events, whether other people are present or not and what the relationship between them might be. For features such as turning lights on and off, the system would not need any triggers at all and would use various sensors, such as motion sensors or Kinect and Playstation Move cameras. In the absence of any of these inputs, the system falls back to traditional means of control, such as remotes or apps. The operating modes include specific motifs and content, such as photos, videos, calendars, social notifications, etc, categorized and prioritized, according to the specific situation. This implies a certain level of communication between the system and the wearable device.

The interaction would look like this:

    - the system is detecting movement and sends a query signal.

    - the device is answering with an identification signal, if authorized, and queries the system for the current status.

    - the system respond with the current status, i.e. " there is nobody else in the room" and/ or "no scheduled tasks"

    - the device would then send a series of commands which can populate the space with pertinent information, lighting setting, photos, notifications, music, entertainment settings such as a specific type of TV programming, etc.

    - if the current status changes, the system notifies the device and the device responds with new commands.

Even though the levels of customization and possible scenarios are virtually unlimited, we have to remember this is a shared family space and in such case, the system will primarily be directed to distinguish private and public modes (when guests are present) as well as children and adult modes. I also believe the most immediate and biggest impact will relate directly to multi-media capabilities and entertainment functions, because these are major driving forces behind the technologies present in our living rooms.

### 4.3. Entertainment considerations.

[…]The optimal next-generation entertainment experience cannot be created just by innovation in content creation or distribution or playback devices. The best results stem from an end-to-end integrated approach across all three to create a single unified ecosystem. (Geller, 2015)

I think that Geller is right, and today, every system incapable of delivering the expected results will immediately require an updated design, even at substantial redevelopment costs, because profits do not come from local or national markets, but rather every product or service can be marketed and sold almost anywhere in the world. And with that comes an unprecedented economy of scale, in all areas mentioned by Geller, content creation, distribution and playback devices. I have taken into consideration these three areas in my concept, but in the reverse order, starting with consumers and ending with content creators.

In a previous chapter I described the decline of TV industry as we know it, going back to that core of news and live events, its "lifeblood and magic" as CBS president Frank Stanton called it (William, 1998). I think this process will sustain the old infrastructure for another 10, maybe 15 years, until broadband services become available worldwide and more affordable. Broadband plays a key role in this space (the living room) and the affordability issue is present even in developed countries such as the US, where "in the most affluent sectors, 80 to 90 percent of households have internet at home. In the regions with the lowest median income, only about 50 percent do" (Dzieza, 2015). This is 2013 data and these numbers may have improved, but it does raise the question whether on top of the broadband cost these households would pay for subscription based streaming services like Netflix, Hulu, etc.

Virtually every playback device on the market today is a front to a specific platform, which limits user choices. The CNET review of the Amazon Fire TV mentions that "the user interface strongly favors Amazon Instant content over other services, and the voice-search feature doesn't comb through Netflix or most other non-Amazon apps" (Moskovciak, 2014). Apple TV and Google Chromecast are similarly aligned with their respective stores. Roku 2 is an exception, its "search is the best on the market, hitting 17 services and arranging results by price" (Katzmaier, 2015). There are other options, like Intel's Compute Stick, a mini computer capable of running Windows 8.1 and able to access services provided by Apple, Amazon or Google. But all these devices, with all the backing from some of the most powerful companies in the world, Google with a 440.04 Billion USD market capitalization, Apple with 629.47 Billion USD and Amazon with 249.07 Billion USD (as of October 2015) fall short of a simple piece of software, Popcorn Time (which, most likely, could run on any of them). It does not require any infrastructure, any networks deployed or any maintenance. This simple piece of software can revolutionize content distribution and content creation overnight. I believe that a legal version of Popcorn Time can be readily deployed as an app on any hardware by any of the companies mentioned above.

Crawford calls Popcorn Time, "the straw that broke the movie industry camel's back" and not because of the loss of revenue associated with piracy but because "it exposes the draconian and artificial hold copyright holders have over content, which is designed to maximize their profits to the detriment of consumers' interests" (Crawford, 2015). He claims that Netflix's US catalogue available for streaming allows access to about 60,000 titles, while the other regional markets "have to make do with a vastly inferior choice", for example UK, a market with access to only about 10,000 titles. (Crawford, 2015). He concludes: "the result is that ordinary members of the public who are happy to pay for content are denied access to the movies and TV shows they want to watch" (Crawford, 2015).

I think content distribution companies underestimate the number of people willing to pay for content. A survey conducted by consumer advocate Choice showed that in only six months, "the arrival in Australia of streaming services like Netflix and Stan […] has resulted in a drop of about 25 per cent in the rate of piracy" (Moncrief, 2015). I believe that a cost efficient device, similar to Google's $35 Chromecast, running a custom version of Popcorn Time can be jointly developed by major content providers and used as a base for a monthly subscription package. This would enable production companies and studios to distribute content directly to consumers, without the need for streaming infrastructure. A device with limited storage capacity, would also provide a very efficient way to control the number of copies available to users. Using a custom filesystem with a limited capacity or even an outdated and unsupported filesystem of around 4 Gigabytes, one could hold no more than 3 high definition or 5 standard definition full length feature films at any given time. I think this should help alleviate the copyright holders' concerns about illegal use of their content, while offloading much of the infrastructure costs to their users.

The simplicity and efficiency of such system is that it automatically scales with demand: as more and more people request certain content, they also become the seeds to spread the content further. As new content is requested from the central facility, older content is purged from the user's cache and the cycle resets. Simple swarm statistics available in real-time from monitoring the number of seeds, would provide a very simple yet efficient tool to gauge which content is consumed and form a base for profit sharing. I think this earnings model would provide a huge incentive for content creators to consistently deliver relevant content and would provide a very direct and accurate "feel" for their customers. It would also level the playing field, giving large and small contributors the same footing in front of their consumer. The two most important hurdles for this service would be to finally acknowledge the price that customers are willing to pay, and to make the same content available in all areas serviced. These are key to unlocking the full potential of this market and, I think, the main reasons behind piracy today.

I mentioned Netflix in the previous section because they are the only company with revenues coming exclusively from distributing content. Apple, Amazon, Google, can use other services or products to offset costs and this might play a factor in setting up their prices. And Netflix is setting up a solid benchmark for its competitors. I believe that a joint effort from content providers, maintaining a Netflix level of service at their price point, would eliminate piracy altogether, or at the very least reducing it to an insignificant level. I suspect that such system will be initially deployed by small and medium sized-content providers, maybe as a regional initiative or by specialized groups, such as the porn industry.

### 4.4. Gaming / virtual reality considerations

In a relatively short period of time, running on faster and cheaper hardware, computer games evolved in complexity and migrated to virtually all digital platforms. As of 2013, this is now a 93 billion dollar market (Gartner Inc., 2013). Bigfishgames.com claims that "over 59% of Americans play games", approximately "150 million people spread over a vast variety of backgrounds, ages, genders, socioeconomic statuses" (Lofgren, 2015). "29% of gamers are under 18", almost as much as the 50+ years olds at 27%, while the 35-44 year olds were the biggest spenders, averaging $6 per person every month (Lofgren, 2015 & Bigfishgames.com, 2015). This is indeed a universal phenomenon and I think that the living room of the future should certainly support any type of gaming platform or device.

However, today the gaming industry is one of the most coercive and least consumer-friendly industries. If we were to apply the same restrictions to television for example, one would have to buy more than one set to view all the channels. In fact, he or she would need one for every major platform: Nintendo, Microsoft and Sony. Each platform thrives on exclusivity of certain titles and in turn, this drives hardware sales. I cannot believe that after the Betacam -VHS war, the Blu-ray vs. HD DVD war, we are still forcing consumers to make unnecessary choices. In the meantime, while Nintendo, Microsoft and Sony are trying to sell boxes, the revenue from mobile gaming "is expected to overtake console gaming" in 2015, with an estimated $25 billion in sales for 2014, a 42% increase over 2013 (Lofgren, 2015). And while both Microsoft and Sony have announced VR sets coming to their platforms in 2016, Samsung is already selling the Gear VR headset alongside their smartphones. Scott Stein from CNET reviewed the headset and concluded: "I never thought I'd see a technology that had that same effect as early Lumiere films, but VR is definitely it." (Stein, 2014).

The scary part for Microsoft and Sony should not be that Samsung already has a phone capable of driving a VR set, but rather that the Android platform has nearly 400,000 developers (Michaeli, 2015). Apple will release a new Apple TV later this year, capable of playing games and a worthy competitor to Nintendo's Wii and Sony's Playstation TV. I think this increased pressure will drive the big three, Nintendo, Microsoft and Sony, to make some bold moves, including opening up their platforms to each other's exclusive games and to other apps. Nintendo has already announced that it will start to make games available on smartphones (Broussard, 2015).

All three gaming systems are capable of streaming video from various sources, and 2 of them also feature voice control and cameras. I firmly believe that the next generation of consoles will be anything but dedicated gaming consoles. In addition to playing games from any store or platform, be capable of supporting Virtual Reality and probably holographic applications, they will become the main connecting hub, providing a common platform for all other devices in our homes. They will support and integrate new functions, like home security, environment and health monitoring, be able to seamlessly manage educational and work-related tasks across multiple devices and platforms. This would also make sense for the 74% of K-8 teachers who use digital games in the classroom, reporting that "video games increase motivation and engagement in their students" (Lofgren, 2015). And beyond the intrinsic educational value, it seems that "a student can exhibit a growth mindset when he or she plays an educational game, struggles with a level, tries again, and again, and then improves their performance to reach the game's goal" (Luna-Lucero, 2014).

So games are good for students on many levels, yet its almost impossible for one to come home from school and continue to learn through play. That is because platforms, both hardware and software are not design to be compatible with each other. I think the current system where hardware and software manufacturers are holding customers hostage, is outdated and unfair. I believe that anyone should be able to access any hardware with a simple and secure set of credentials, conveniently stored on a smartphone or other portable device and be able to use the software he/she owns and the services that he/she has subscribed to, roaming from one box to another, able to continue its work and play without interruption. And because the hardware is separate from the data storage device, there are no privacy and security issues. Again, fairness should play the most important role, just as much as in the content distribution system.

### 4.5. Why seek alternative models?

When it launched in July 2008, Apple's App Store was not the first or the biggest app store. "Microsoft was touting more than 18,000 applications for its Windows Mobile operating system […] while Palm was claiming 30,000 active software developers" (Ranger, 2015). Fast forward to June 2014, and the App Store hosts 1.4 million apps (up from 500 when it launched), available in 155 countries (Ranger, 2015). Ranger also claims that it has the highest revenues among its competitors, even though Google Play, its Android rival, has more apps and more developers working on the platform (Ranger, 2015). The single most important reason why developers have scrambled to make apps for Apple's devices was their profit sharing system: 70% of the proceeds would go to developers and Apple keeps the other 30%. Steve Jobs, was quoted in this article, saying that this was the best deal available to developers at the time (Ranger, 2015).

According to Market capitalization figures (sourced from wikinvest.com) Apples value grew from $126.12B in 2008, to $3,175,75B in 2014, but this has very little to do with the App Store's profit, estimated to be around $10B (Ranger, 2015). Their profits come from building devices to run this software, most notably the iPhone. According to Blodget, even as far as 2012, the iPhone was already the most profitable product in the world, "so profitable that it generates more profit than just about any other company on earth, not just product" (Blodget, 2012). But the iPhone itself, the device, does not warrant such success, in fact, as Mack points out, "other flagship phones have had better specs on paper than the iPhone for years now" (Mack, 2015). Tshe one thing that differentiates the iPhone from its competitors is the App Store and the developers who sell their software in it. They have become the most valuable "asset" for Apple. This is what van Riel calls a service constellation. He argues that:

> Even when consumers consider buying and using a single service, they (implicitly or explicitly) take into account the actual or future existence of other services that (positively or negatively) affect the value of what they, at that point in time, consider to be the focal service. (van Riel et al., 2013)

Van Riel et al., concedes that "the concept of service constellations makes service innovation decision-making more complex than focusing on the development of individual services in isolation" but they maintain, referencing Cooper and Kleinschmidt, Langerak et al., Narver et al., that "this approach also takes the consumer perspective more seriously and thus makes organizations more market-oriented" (van Riel et al., 2013). I think his observation strengthens my previous arguments about treating customers fairly, both in content distribution and as consumers of hardware/software products. Van Riel acknowledges that innovation within the service constellation model requires a different approach and highlights some of the differences:

Traditional versus service constellation approach, (van Riel et al., 2014)

|  | Individual service innovation approach | Service constellation approach |
| --- | --- | --- |
| Legal structure | Sole ownership | Shared or distributed ownership |
| Organizational structure | Hierarchical | Decentralized, rhizomatic |
| Knowledge structure | One good idea<br>Closed/patented | Network of ideas<br>Open source/shared |
| Social structure | Local interest optimization<br>Isolation<br>Competition | Global optimization<br>Community<br>Co-operation |
| Theoretical perspectives | Economic theories<br>Transaction cost economics<br>Psychological perspective<br>Deterministic | Complexity theory Organizational ecology Sociological perspective Probabilistic |
| Creativity | Lightning bolt (Newton) | Collaborative innovation |
| Models of science | Mechanistic/deliberate | Emergent Distributed/dispersed |
| Power/decision-making | Centralized Rational/analytic | Complex/integrative |
| Challenges | Finding a brilliant idea<br>Enclosing property | Selecting good ideas Distributing value fairly |
| Success metrics | Market share<br>Competitive advantage<br>Profit | Survival/sustainability<br>Status/reputation/credibility/trustworthiness<br>Centrality of the system, dependence on others |

Van Riel's suggests that these models are not mutually exclusive, but rather they overlap. And there are some examples of that, Apple being one of them. As a public company, they operate consistently with the Individual service innovation approach, but there are certain aspects of their business where the constellation approach has been deeply implemented, like the App Store. The most interesting aspect of this symbiosis is that the company as a whole, is benefiting in a huge way from it. We can learn a lot from this approach and can expand the concept to include any number of companies and products. Again, I think that one should be able to buy any type of device or hardware and know that he or she can run any type of software or service from any source with it. This would be truly liberating for all consumers, and would essentially eliminate a great deal of anxiety and uncertainty usually associated with purchasing new technology and related services. I am not implying that we should stop manufacturing the variety of devices that we are right now, rather I am suggesting that we should open up these platforms and adopt open standards across the board. A notable example is the Wi-Fi Alliance, comprising about 600 companies committed to produce compatible hardware, software and services. It hasn't always been a smooth operation, with patent infringement lawsuits worth more that $1 billion, but it significantly alleviated consumer concerns and consequently improved the adoption rate of this technology (Moses, 2010).

It's basically the same concept I have suggested for content distribution, where I proposed that content creators should pool their products together. Here, we have software developers and hardware manufacturers selling compatible products and services. While it may seem that we already have such model in Google's Android platform, supported by Google's Play Store and serviced by a large number of hardware manufacturers, there is a major caveat that it's holding this whole service constellation back. And that is what happens when things do not go smoothly, which is often the case. Whenever something does not work as intended in Apple's service constellation that issue will be dealt with by Apple. On the Android platform that responsibility is shared between Google, device manufacturers and app developers and this is a major flaw, ultimately impacting ngatively each other and the platform as a whole.

I think we can much better than this. This is what fuelled the proposal of such model. From architectural elements, to content availability, to entertainment capabilities, to educational and work-related activities, this model is built around choice and flexibility for consumers. The service constellation model can be applied to almost any industry and finding ways to distribute profits in a fair way, without asking consumers to make this choice should be a priority. I think above all, consumers value the experience, the end result and this is the only thing they should be asked to choose, not manufacturers, devices and services.

## 5. Conclusion

I think the living room is about to be transformed, morphed into a family friendly, entertainment and communication hub, a direct consequence of how generation Y is appropriating various technologies. Historically, computing technologies followed three major directions, increasing productivity, enhancing communications and developing entertainment platforms, and all three streams developed around very broad market demands. Over the years, these general demands crystallized into narrower applications and the large grey boxes hidden under desks transformed into a multitude of specialized devices targeting every individual's particular needs.

Generation Y, the driving force behind this transformation, as they age and move into their own family environments will influence the development of new technologies but from a different perspective. Privacy and security concerns will become major issues for them as they transition from their earlier social lives to a more private, family oriented setting. The living room, as private and public space, will also have to balance their digital lives, acquiring the necessary technology to support this dimension and become the gateway between their private and public digital identities.

The introduction of ubiquitous computing, independent of device, location, and format, already has measurable effects on our society, most notably in communicating with each other. It also changed the way we control computers, with voice and gesture inputs becoming just as efficient as keyboards and other peripherals. But our homes have been relatively slow in adopting these technologies. We see early adopters trying out different devices, (Merriman, 2015) and we see a tremendous activity from companies in this space: Google is actively developing an operating system for IOT devices - project Brillo (Page, 2015), Apple is quickly expanding its HomeKit (Etherington, 2015), Intel is acquiring expertise to "fuel growth in the cloud data centre and Internet of Things (Page, 2015). And my favourite, "weekend ruiner and flatpack assembly company Ikea is working with Samsung to release furniture for the home or workplace that can wirelessly charge your gadgetry gewgaws" (Neal, 2015).

However, a dominant platform or even an industry-wide, agreed upon set of standards has not emerged yet and there are numerous privacy and security concerns around ubiquitous computing and the Internet of Things (Internet crime Complaint Center, 2015). This has a dual effect on this industry: it fuels innovation by constantly seeking out a better solution and in the same time, it gathers momentum, leading up to the inevitable breakthrough or failure.

I think this evolution will be guided by new entertainment platforms further timulating the adoption of devices and technologies. These technologies have long superseded the one way flow of information that radio and television brought into our homes. The latest wave of technological powerhouses are built on "observing and fusing publicly available data, such as web search queries, blogs, micro-blogs, internet traffic, financial markets, traffic webcams, Wikipedia edits, and so forth" which is used to "anticipate events such as disease outbreaks, financial and political crises, economic instability, resource shortages, and responses to natural disasters. (Mordini, 2014). I think we are witnessing a move towards individualization and personalization of software and services, similar to the hardware transformation from grey boxes into the very specialized devices we have today. Indeed, as a 2009 advert for the iPhone 3G points out, "there is an app for just about everything" (Apple, 2009). Consequently, this bidirectional data flow enabled with these personalized apps is not so public anymore and I think we should be able to decide whether that data exists or not and if it does who will use it and how. Ultimately, this would be an agreement similar to the one we had with commercial television, programming in exchange for commercials, and it needs to be just as transparent and simple.

I think it's important to revisit and consider the changes which have occurred over the years with other technologies, like television, which had a long lasting, trans-generational impact, on almost every social, economical and political aspect of our lives. And finally I think we need to reframe gaming as a concept, redefine the notion of computer games, in order to better understand the applications and social implications of this phenomenon.

I have attempted to anticipate some of the changes which I think are imminent in this space and provide sensible, realistic solutions to some of the issues we face today. I can only hope that some of these ideas will serve as inspiration to others when researching or considering technological innovation in our homes and in our living rooms in particular.